\documentclass[11pt]{article}

\usepackage{geometry} 
\usepackage{titlesec}
\titlespacing*{\section}{0pt}{0.7\baselineskip}{0\baselineskip}
\titlespacing*{\subsection}{0pt}{0.7\baselineskip}{0\baselineskip}
\titlespacing*{\subsubsection}{0pt}{0.7\baselineskip}{0\baselineskip}

\usepackage[utf8]{inputenc}
\usepackage{lmodern}

\usepackage{authblk}

\usepackage{xcolor}

\usepackage{amsmath}
\usepackage{amssymb} 
\usepackage{mathtools}	
\usepackage{bm}
\usepackage[makeroom]{cancel}

\usepackage{siunitx}

\usepackage[english]{babel}

\usepackage{graphicx}
	\graphicspath{{eps_fig/}}
\usepackage{subcaption}
\usepackage{setspace}  
\usepackage{epsfig}	

\usepackage{multirow}
\usepackage{booktabs} 
\usepackage{array}
\newcolumntype{C}[1]{>{\centering\let\newline\\\arraybackslash\hspace{0pt}}m{#1}} 
\newcolumntype{P}[1]{>{\raggedright\arraybackslash}p{#1}} 

\usepackage{nomencl}
\makenomenclature

\usepackage{cite}

\usepackage[colorlinks=true,
	linkcolor=blue,
	citecolor=blue,
	urlcolor=blue,
	filecolor=blue,
	pdfstartview=FitH]{hyperref}
	
\usepackage{fancyvrb} 
\usepackage{soul} 
\usepackage{enumitem}

\usepackage[bottom]{footmisc} 

\usepackage{lineno}

\definecolor{tgreen}{rgb}{0.0, 0.5, 0.0}

\begin{document}

\title{\vspace{-2.0cm}\textbf{Topology synthesis of a 3-kink contact-aided compliant switch} }
\date{}
\author[$\dagger$]{B V S Nagendra Reddy}		
\author[ ]{Anupam Saxena}
\affil[ ]{\normalsize Department of Mechanical Engineering, IIT Kanpur, India 208016.}
\affil[$\dagger$]{\small Corresponding author: \url{bvsnagendrareddy@gmail.com}, \url{bvsnr@iitk.ac.in} }

\maketitle
\begin{abstract}
\noindent
A topology synthesis approach to design 2D Contact-aided Compliant Mechanisms (CCMs) to trace output paths with three or more kinks is presented. Synthesis process uses three different types of external, rigid contact surfaces — circular, elliptical and rectangular — which in combination, offer intricate local curvatures that CCMs can benefit from, to deliver desired, complex output characteristics. A network of line elements is employed to generate topologies. A set of circular subregions is laid over this network, and external contact surfaces are generated within each subregion. Both, discrete and continuous design variables are employed — the former set controls the CCM topology, appearance and type of external contact surfaces, whereas the latter set governs shapes and sizes of the CCM constituents, and sizes of contact surfaces. All contact types are permitted with contact modeling made significantly easier through identification of outer and inner loops. Line topologies are fleshed out via a user-defined number of quadrilateral elements along lateral and longitudinal directions. Candidate CCM designs are carefully preprocessed before analysis via a commercial software and evolution using a stochastic search. The process is exemplified via a contact-aided, 3-kink mechanical switch which is thoroughly analysed in presence of friction and wear.

\end{abstract}


\section{Introduction}

\label{sec:introduction}

Contact-aided compliant mechanisms (CCMs) are special cases of compliant mechanisms (CMs) that undergo contact while transferring force/motion by means of (large) elastic deformation of its members. Compliant mechanisms offer advantages of zero backlash, high repeatability, assembly-free manufacturing amongst many others, all possible due to absence of rigid-body joints. Consequently, compliant mechanisms readily find applications in areas like electro-mechanical devices, e.g., \cite{masters2005three}, medical equipment, e.g., \cite{frecker2005design} and mechanical devices, e.g., \cite{yin2004optimal}. Contact-aided compliant mechanisms inherit most advantages of compliant mechanisms, and with contact permitted between their constituents and external surfaces and/or between members themselves, CCMs can be designed to deliver intricate mechanical characteristics. CCMs have many applications in compliant joints \cite{cannon2005compliant,moon2007bio}, surgical tools \cite{eastwood2018design,halverson2008flexure}, mechanical devices \cite{howell2001compliant,saxena2013contact, howell2013compliant,song2019new} and  flapping wing mechanisms\cite{tummala2014design,calogero2018tuning} and offer immense potential yet to be tapped. Many design methods exist for compliant mechanisms that could be broadly classified into {\it pseudo-rigid-body model} and {\it structural optimization} approaches \cite{howell1994method,howell1996loop,howell2001compliant,edwards2001pseudo,jensen2003identification,masters2005three,sonmez2008compliant,su2009pseudorigid,howell2013compliant}. However, few address a generic design procedure for Contact-aided compliant mechanisms. Mankame and Ananthasuresh\cite{mankame2004topology} are the first to introduce Contact-aided compliant mechanisms to impart special displacement characteristics to CMs.  They demonstrate synthesis of CCMs by allowing their members to interact with pre-identified external surfaces in the {\it mutual contact mode} \cite{mankame2007synthesis} to trace non-smooth ($ C^{0} $) paths. Contact between members of the continuum also contributes in CCMs tracing non-smooth paths. Such, \textit{self contact} CCMs are designed in \cite{Reddy2012systematic}, wherein contact location information is not available \textit{a-priori}, rather determined systematically. Using this methodology, a displacement delimited gripper is designed in \cite{saxena2013contact}. Kumar et al. \cite{kumar2019computational} design CCMs for paths with mostly single kinks wherein the design region is represented with hexagonal tessellation, material mask overlay strategy is employed for topology design, and both self, and mutual contact modes are considered assuming frictionless contact. Yin and Lee \cite{lan2008analytical} propose an analytical contact model to design compliant fingers. Mehta et al. \cite{mehta2009stress} demonstrate increase in stretching capacity of contact-aided cellular structures due to the induced stress relief. Jin et al. \cite{jin2020large,jin2020cprbm} propose methodologies to analyze CCMs undergoing mutual contact, beam-to-beam contact and beam-to-rigid contacts. 

\section{Aim, novelty, scope and organization}

\label{sec:aim}

The objective, and novelty of this work is systematic continuum design of large deformation CCMs that trace non-smooth paths with multiple (three or more) kinks. To accomplish this, we propose using external surfaces of three different kinds -- circular, elliptical and rectangular. These surfaces, individually or in combination, can offer a variety of local curvatures that CCMs can interact with and benefit from, in delivering the desired intricate characteristics. The methodology proposed herein is significantly improved compared to those in \cite{Reddy2012systematic} and \cite{kumar2019computational}. In \cite{Reddy2012systematic}, only self-contact between members is considered. In \cite{kumar2019computational}, CCMs are synthesized using hexagonal cells with both, self and mutual contact modes, however, rigid external surfaces used are circular in shape. We permit both, self and mutual contact modes in our designs. Compared to \cite{Reddy2012systematic}, (i) contact surfaces are identified efficiently by modeling each as a closed loop. (ii) Many 2D elements are used across frame members to model the latter, thereby addressing mesh independence, a feature absent in both \cite{Reddy2012systematic} and \cite{kumar2019computational} wherein a single element across member width is employed. (iii) Contact interactions between members and junctions and those with free ends of dangling members are all considered. (iv) Care is taken to ensure absence of external surfaces close to the output port, to avoid trivial CCM solutions. Contact is assumed frictionless and external surfaces are rigid. Design generation, evaluation and evolution is performed via \texttt{MATLAB\textsuperscript{TM}} while all intermediate/final designs are analysed using a commercial software, \texttt{ABAQUS\textsuperscript{TM}} which, notwithstanding computational time, offers generic analysis capabilities.  

The outline of the paper is as follows. In Sec. \ref{sec:synthesis}, design parameters, methodology for mesh generation of the continuum and that for generating and modeling  potential contact surfaces are presented. While a commercial software is used for design evaluation, large deformation friction-free contact analysis is summarized in Sec. \ref{Large_Deformation_contact}. Sec. \ref{sec:objective} discusses the search algorithm. Synthesis of a 3-kink Contact-aided compliant switch is presented in Sec. \ref{sec:examples}, while in Sec. \ref{sec:Discussion}, the 3-kink switch is analysed further for effects of friction and wear. Aspects covering computational efficiency, effect of mesh size on final solution and search algorithm are also discussed, and finally, conclusions are drawn in Sec. \ref{sec:conclusion}.
	
\section{Methodology}
\label{sec:synthesis}
Procedure for generation of a candidate CCM design is described below.
	\subsection{Generation of candidate solution}	
	\label{sec:design_setup}
\begin{figure}[h!]
			\centering				
				\includegraphics[scale=1]{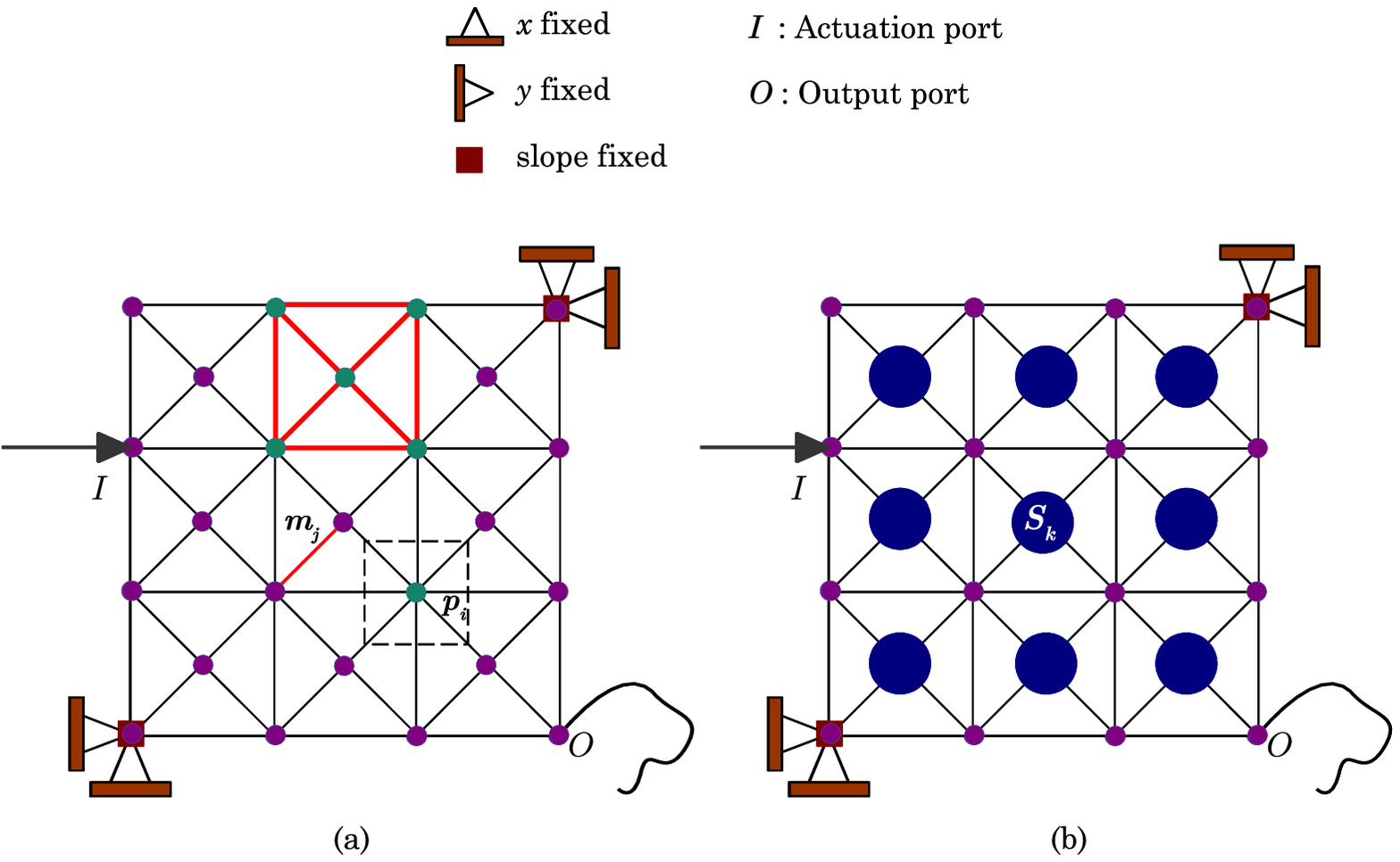}
				\caption{Design domain without (a) and with (b) external contact surfaces. Set of \textit{eight} \textit{red}  lines in (a) represent one   block. \textit{Blue} circles in (b) represent initial placement of external contact surfaces. ${I} $ represents the input/actuation port, while, ${O} $ represents the output port.}
				\label{fig:Design_Domain}
		\end{figure}

		The initial design domain comprises of blocks of either square or rectangular shape (Fig. \ref{fig:Design_Domain}a) over which external objects are placed (Fig. \ref{fig:Design_Domain}b). Each block consists of eight members connected with five vertices.  External objects are initially assumed circular in shape and are  distributed  uniformly over the domain space (\textit{nine} external contact surfaces are shown in Fig. \ref{fig:Design_Domain}b).
				Each member $m_j$ in the initial domain is associated with both, continuous and discrete variables \cite{Reddy2012systematic}. Continuous variables are varied between the upper and lower bounds specified by the user. Discrete variables take the binary value of either 0 or 1. A different set of continuous and discrete variables define each external surface $S_k$. Design variables for both members and external surfaces are detailed hereunder.
		
\subsection{Design variables for members}  Presence of each member $m_j$ in a candidate design is decided by the continuum topology choice variable $v_{m_j}^c$, that assumes a value of either $0$ or $1$. $v_{m_j}^c=0$ indicates  absence of the $j^{th}$ member in the topology, while, $v_{m_j}^c=1$ denotes its presence. If required, as explained in Sec. \ref{sec:Design_Preprocessing}, all members intersecting with any external surface of the candidate design are filtered out by assigning $v_{m_j}^c$ to $0$. Each member, whether present or absent, is modeled as a Hermite cubic curve with end slopes $v_{m_j}^{T_1}$ and $v_{m_j}^{T_2}$, both continuous variables. Each end slope varies between the user specified limits [$v^{ {T}}_L, v^{ {T}}_U$]. The in-plane width $v_{m_j}^w$, a continuous variable for each member $m_j$, varies between the lower and upper bounds [$v_L^{w},v_U^w$]. The out-of-plane thickness $v^{th}$ is the same for all members of the continuum and varies between $v_L^{th}$ and $v_U^{th}$. Also, each vertex $i$ of the domain, with coordinates [$v_{p_i}^x, v_{p_i}^y$], can move freely within a rectangular space controlled by the user specified limits [$v_L^x, v_U^x$] and [$v_L^y, v_U^y$]. As suggested by Borrvall and Petersson \cite{borrvall2001topology}, a continuous variable $v_{I}^F$ representing the magnitude of the input force acting on node $ I $ is considered. $v_{I}^F$ is varied between the user specified limits [$v_{L}^F, v_{U}^F$] like other continuous variables.
		
\subsection{Design variables for external surfaces} Similar to the discrete choice variable $v_{m_j}^c$ for members, $v_{s_k}^c$ represents the contact topology choice variable for the $k^{th}$ external surface ($S_k$). $v_{s_k}^c$ can be either $0$ or $1$ based on its presence ($v_{s_k}^c=1$) or absence ($v_{s_k}^c=0$).  In addition to the topology contact choice variable ($v_{s_k}^c$), \textit{one} discrete ($v_{s_k}^{sh}$) and \textit{six} continuous ($v_{s_k}^x, v_{s_k}^y, v_{s_k}^R, v_{s_k}^{f1}, v_{s_k}^{f2}, v_{s_k}^{\theta}$) variables, define each external surface. The discrete shape variable $v_{s_k}^{sh}$ possesses information about the shape of the $k^{th}$ external surface. Each external surface is either circular, elliptical or rectangular in shape. Accordingly, $v_{s_k}^{sh}$ is assigned with $1$ for circular, $2$ for elliptical, and $3$ for rectangular shapes. Variables [$v_{s_k}^x, v_{s_k}^y$] represent position of the center of $k^{th}$ external surface. Lower and upper limits of the centroid of the $k^{th}$ external surface are set to the size of design domain, i.e., the centroid of $k^{th}$ external surface can lie anywhere within the design domain. $v_{s_k}^R$ indicates the radius of the bounding circle that encloses the $k^{th}$ external surface. $v_{s_k}^R$ varies between the lower and upper bounds [$v_L^R, v_U^R$] as specified by the user. $v_{s_k}^{f1}$ and $v_{s_k}^{f2}$ are continuous factors, varying between 0 and 1, that help determine the size of $k^{th}$ external surface when combined with the radius of the enclosing circle, $v_{s_k}^R$.	
	
\noindent For circular external surfaces, radius of the $k^{th}$ external surface  is	
		\begin{align}
			R_{s_k} =  \frac{v^{f1}_{s_k} v^R_{s_k} + v^{f2}_{s_k} v^R_{s_k}}{2} 	
		\end{align}

\begin{figure}[h!]
			\centering			
			\includegraphics[scale=1]{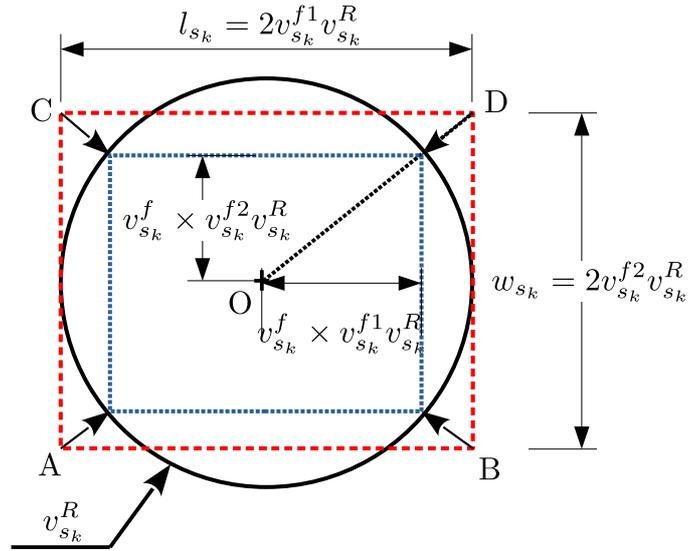}
			\caption{Resizing of external rectangular contact surface (dashed \textit{red}) whose corners A, B, C and D are initially outside the bounding circle (\textit{black}) of radius $v^R_{s_k}$. \textit{Blue} dotted lines represent the resized surface.}
			\label{fig:Resize_ext_body}
		\end{figure}

\noindent For elliptical external surfaces, semi-major axis of the $k^{th}$ external surface is $a_{s_k}=v^{f1}_{s_k} v^R_{s_k}$ and the semi-minor axis is $b_{s_k}=v^{f2}_{s_k} v^R_{s_k}$. For rectangular external surfaces, length of the $k^{th}$ external surface is $l_{s_k}=2 v^{f1}_{s_k} v^R_{s_k}$ and width is $b_{s_k}=2v^{f2}_{s_k} v^R_{s_k}$. However, if corners of a rectangular surface are not completely within the bounding circle, variables $v_{s_k}^{f1}, v_{s_k}^{f2}$ are scaled down further by multiplying with a parameter $ v^f_{s_k}$, where
		\begin{align}
			 v^f_{s_k} =  \frac{v^R_{s_k}}{\sqrt{(v^{f1}_{s_k} v^R_{s_k})^2 + (v^{f2}_{s_k} v^R_{s_k})^2}} 
		\end{align}
		so that corners of the external rectangular surface always lie within the bounding circle as shown in Fig. \ref{fig:Resize_ext_body}. This, though not necessary, is done to maintain consistency in having all external surfaces lie within their bounding circles. For elliptical and circular external contact surfaces, resizing is not applicable since they always lie within the respective bounding circles.  Finally, orientation of the external surface is defined by $v_{s_k}^{\theta}$ (for a circular surface, this variable is redundant), that takes the lower and upper bounds of $0$ and $\pi$ respectively.
		
		\subsection{Stage II candidate solution}
		\begin{figure}[h!]
			\begin{subfigure}{0.48 \textwidth}
				\centering
				\includegraphics[scale=1]{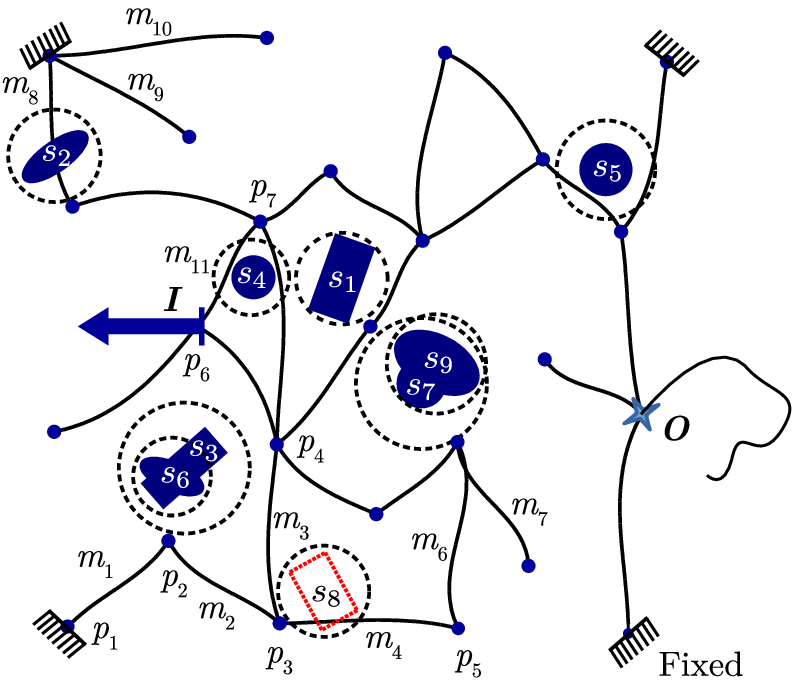}
				\caption{}
				\label{fig:Candidate_solution}
			\end{subfigure}	\hfill
			\begin{subfigure}{0.48 \textwidth}
				\centering
				\includegraphics[scale=1]{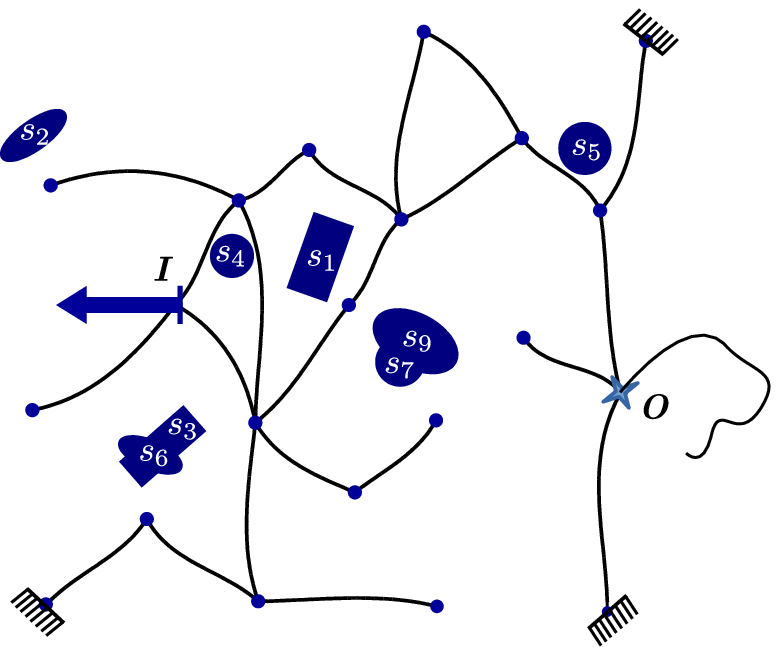}
				\caption{}
				\label{fig:Candidate_solution_2}
			\end{subfigure}				
			\caption{(\subref{fig:Candidate_solution}) A candidate solution, with external surfaces, fixed at specified corners. Each external surface with $v_{s_k}^c =1 $ is shown in \textit{blue}. External surface $S_8 $ with $v_{s_k}^c =0 $ is shown with \textit{red} dotted line to exemplify its absence. \textit{Black} dashed lines represent the bounding circles of external surfaces. $ I $ and $ O $ represent the input and output ports respectively. External surfaces $ S_3, S_6 $ and $ S_7, S_9 $ overlap to form complex shaped external surfaces exhibiting more variety in surface curvatures. Surface $ S_2 $ intersects with member $m_8$ resulting in non-consideration of the latter in the analysis. Members $ m_6 \text{ and } m_7$ intersect with each other and are both not considered in further processing. $ p_3 $ represents the common junction of members $ m_2, m_3 $ and $ m_4 $. $p_i$, $i = 1, 2, ..., 7$ are similar junctions. (\subref{fig:Candidate_solution_2}) Candidate solution after ignoring members $ m_j ,  j= 6, 7, 8, 9 \text{ and } 10$. External surfaces are shown without bounding circles. This solution is ready to be \textit{fleshed-out} for further processing and analysis. }
			\label{fig:Candidate_solutions}
		\end{figure}

Given all discrete and continuous variables for both, design members and external surfaces, a candidate solution can be generated. An example candidate solution, as depicted in Fig. \ref{fig:Candidate_solution}, comprises of all members and external surfaces with topology choice variable $v_{m_j}^c =1$ and  $v_{s_k}^c = 1 $ respectively. The absent external surface ($ S_8 $ for which $v_{s_k}^c = 0 $) is shown with \textit{red} dotted line only for representation purpose. Bounding circles of all external surfaces are shown as \textit{black} dashed lines. One observes that the external surface $ S_2 $ intersects with member $ m_8 $. One has a choice of either deleting the member and retaining the surface, or vice versa. The third option is to delete both. We choose to retain more external contact surfaces, for better design choices, and thus, members intersecting with external surfaces are filtered out by assigning the corresponding topology choice variable of members $v_{m_j}^c$ to 0. One notes that deletion of member $ m_8 $ makes members $ m_9 $ and $ m_{10} $ as well as a fixed boundary form a sub-continuum separate from main continuum containing the output port. For simplicity, all sub-continua not connected to the main design are deleted. An option of retaining such sub-continua which are fixed and can interact with the primal design will be explored in future. As members $ m_6 $ and $ m_7 $ intersect, they are also removed by assigning the corresponding topology variables to zero. Candidate design after deleting intersecting members ($ m_6 $ and $ m_7 $), those intersecting with external surfaces ($ m_8 $) and those not connected to the main continuum ($ m_9 \, \mathrm{and} \,m_{10} $) is shown in Fig. \ref{fig:Candidate_solution_2}. It is observed next that the external surface $ S_4 $ is very close to member $ m_{11} $, but does not intersect with it when we consider the continuum skeleton. All such external surfaces that are very close to members are deleted, if found intersecting with a member, after converting the wire mesh to a corresponding quadrilateral mesh as explained in Sec. \ref{sec:Design_Preprocessing}. Stage II candidate design is checked for absence of the input port, output port and fixed vertices. If the input and/or output port is absent and if no fixed vertex is present, the design is penalized. Else, it is preprocessed and prepared for analysis.
					
\subsection{Design preprocessing}
	\label{sec:Design_Preprocessing} 			
		\subsubsection{Mesh conversion from skeleton to 2D elements}
		With wire/skeletal meshes representing an assemblage of beams, contact between only the neutral axes of members can be analysed. For a realistic finite element model, it becomes essential to discretize members with quadrilateral elements to capture the contact behavior accurately. A new and improved \textit{mesh independent} methodology compared to the one in \cite{Reddy2012systematic} is proposed below.
			
		\begin{figure}[h!]
			\captionsetup[subfigure]{position=b}
			\centering					
			\subcaptionbox
			{Discretization of member-$ m_1 $ with quadrilateral elements\label{fig:mesh_1Dto2D}}			    		{\includegraphics[scale=1]{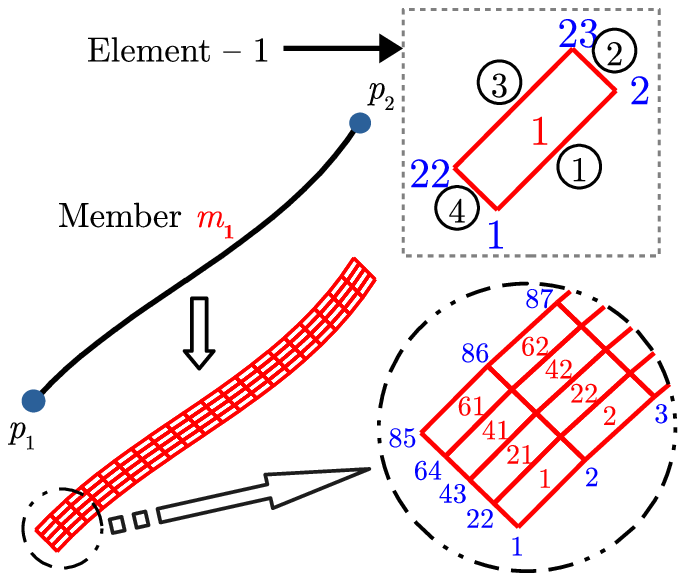}}
			\hfill
			\subcaptionbox
			{Members at the junction without proper connectivity when discretized with quadrilateral elements\label{fig:mesh_disturbed_junction}}
			{\includegraphics[scale=1]{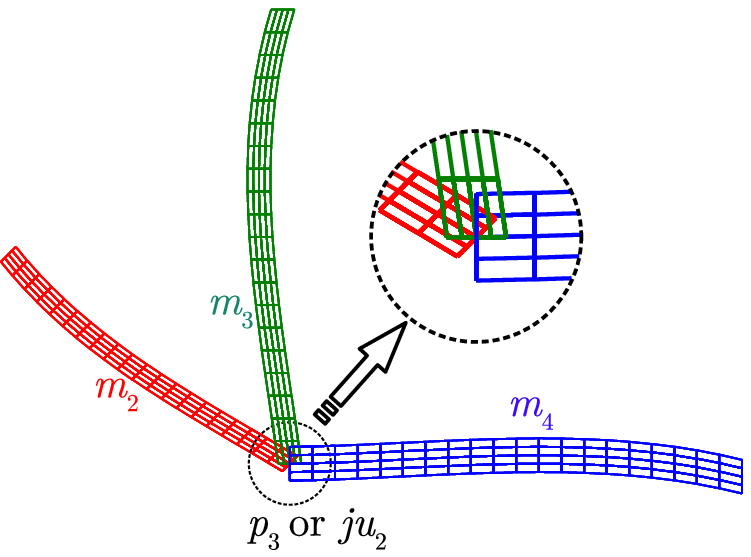}}
			
			\caption{(\subref{fig:mesh_1Dto2D}) Discretization of design members with 4 node quadrilateral elements with node and element numbering. One row of quadrilateral elements along the width of a member (e.g., element numbers 1, 21, 41, 61)  are considered as one \textit{w-set}. A similar row of the quadrilateral elements along the length of the member (e.g., element numbers $ 1, 2,..., 20 $) is referred to as \textit{l-set}. A sample element, with side numbers of the element encircled, is also shown. (\subref{fig:mesh_disturbed_junction}) Design members $ m_2, m_3 $, and $ m_4 $ overlapping at junction $ ju_2 $ located at $ p_3 $, are shown in \textit{red, green}, and \textit{blue} colors respectively}
			\label{fig:mesh}
		\end{figure}
		
\begin{figure}[h!]
			\centering			
			\includegraphics[scale=1]{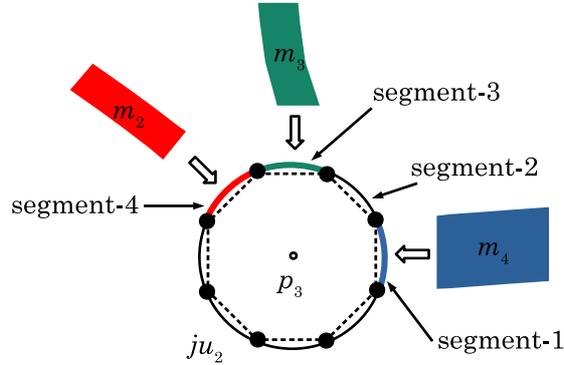}
			\caption{Junction $ ju_2 $ located at vertex-$ p_3 $ (ref: Fig. \ref{fig:Candidate_solution}) modeled as circle of radius $R_{ju}$ based on the side of the imaginary octagon (shown with dashed line) that depends on the in-plane widths of members $ m_2, m_3 $ and $ m_4 $. Members $ m_4, m_3 $ and $ m_2 $ are attached to arc segments 1, 3 and 4 respectively.}
			\label{fig:OctagonalJunction}
		\end{figure}

All curved frame members are divided into $n_{el}$ segments of equal length along the respective arc-lengths. Based on the in-plane width $v_{m_j}^w$ of each member, width of the quadrilateral element corresponding to the $ m_j^{th} $ member is computed as $\frac{v_{m_j}^w}{n_{ew}}$. $ n_{ew} $ indicates the number of elements along the width of each member. Based on the required mesh density, a user can assign any reasonable value to both $n_{el} \text{ and } n_{ew} $. Herein, each member is discretized into 80 quadrilateral elements by choosing $n_{el}=20$ and $n_{ew}=4$. Discretization of member-$ m_1 $ along with node (\textit{blue}) and element (\textit{red}) numbering of selected area is shown in Fig. \ref{fig:mesh_1Dto2D}. A row of quadrilateral elements along the width of a member is referred as a \textit{w-set}. Similarly, one row of quadrilateral elements along the length of member is considered as \textit{l-set}. Each quadrilateral element has four sides  SURF\textsubscript{1}, SURF\textsubscript{2}, SURF\textsubscript{3}, and SURF\textsubscript{4} (numbers encircled on top right in Fig. \ref{fig:mesh_1Dto2D}). Side of any quadrilateral element is represented as $ \mathrm{SURF}^e_d $, where, $ e $ represents the element number and $ d $ represents the side. To specify contact interactions easily between surfaces, all nodes of quadrilateral elements are numbered in sequence such that, for each member, group of SURF\textsubscript{1} edges of all boundary elements along the length represent SURF\textsubscript{1} of the member. Similarly, group of SURF\textsubscript{2} edges  of boundary elements represent SURF\textsubscript{2} of the member and so on. Connectivity at the junctions gets disturbed when the wireframe member is discretized with quadrilateral elements as shown in Fig. \ref{fig:mesh_disturbed_junction}.  To connect members at junctions, each junction, $ ju $, is modeled as a circle of radius $R_{ju}$, and quadrilateral elements (of members) are connected to the peripheral nodes of the circular junction. Each circular junction contains $ 8\times n_{ew} $ peripheral nodes. Also, the periphery of each circular junction is divided into \textit{eight} segments, and each segment is connected to a unique member (if present) of the candidate design (as detailed in the following section). Radius of any junction, $R_{ju}$ is computed based on the maximum in-plane width of all members sharing the junction. For example, radius of junction $ ju_2 $ located at vertex-$ p_3 $ (Fig. \ref{fig:mesh_disturbed_junction}) that  connects members $  m_2, m_3  \text{ and } m_4 $ is given as
		\begin{align}
		R_{2} = & \frac{0.85 \times \text{max }(v^w_{m_2}, v^w_{m_3},v^w_{m_4})}{\sqrt{2- \sqrt{2}} }
		\end{align}
		where $ v^w_{m_2}, v^w_{m_3}, v^w_{m_4} $ are the in-plane widths of members $ m_2, m_3 $ and $ m_4 $ respectively. $ R_{ju} $ is the radius of the circle that circumscribes an imaginary regular octagon of side equal to 85\%\footnote{We chose 85\% after several trials so that the junction doesn't look very big. } of the maximum in-plane width of members sharing the junction $ ju $. Since in the entire domain, maximum \textit{eight} members can share a junction, an imaginary octagonal representation is considered at every junction. Circular junction ($ ju_2 $) along with imaginary octagon (shown with dashed line) located at vertex $ p_3 $ is shown in Fig. \ref{fig:OctagonalJunction}.
		
		\begin{figure}[h!]
			\captionsetup[subfigure]{position=b}
			\centering					
			\subcaptionbox
			{Deletion of elements of \textit{w-set} if the centroid of any element of a set falls inside the circular junction\label{fig:Junction_stage_1}}
			{\includegraphics[scale=1]{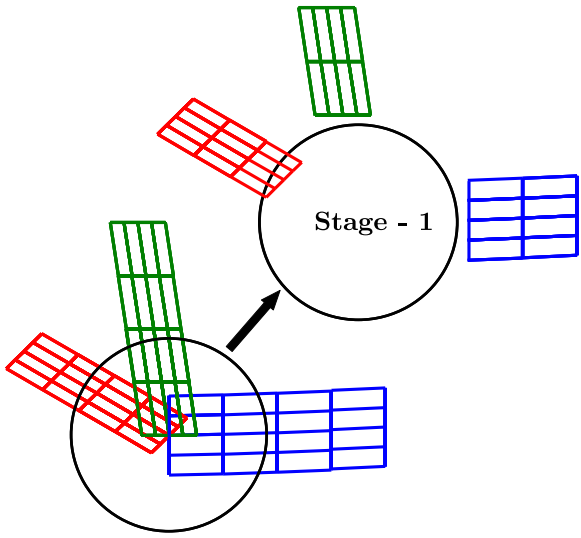}}
			\hfill
			\subcaptionbox
			{Discretization of circular junction and connection of elements of design members with the peripheral nodes of circular junction\label{fig:Junction_stage_2}}
			{\includegraphics[scale=1]{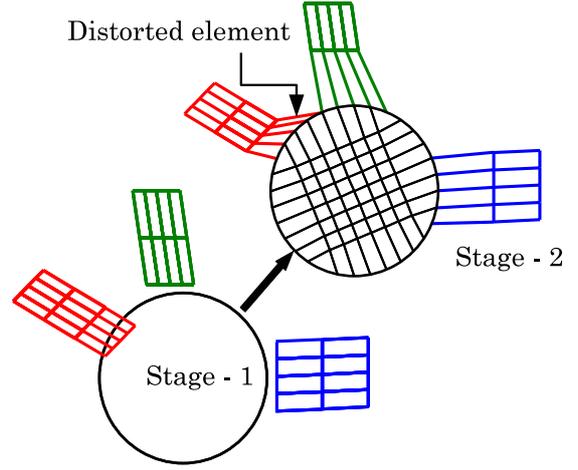}}
			\vspace{\fill}
			\subcaptionbox
			{Repositioning of nodes to reduce distortion of some elements while connecting to junctions\label{fig:Junction_stage_3}}
			{\includegraphics[scale=1]{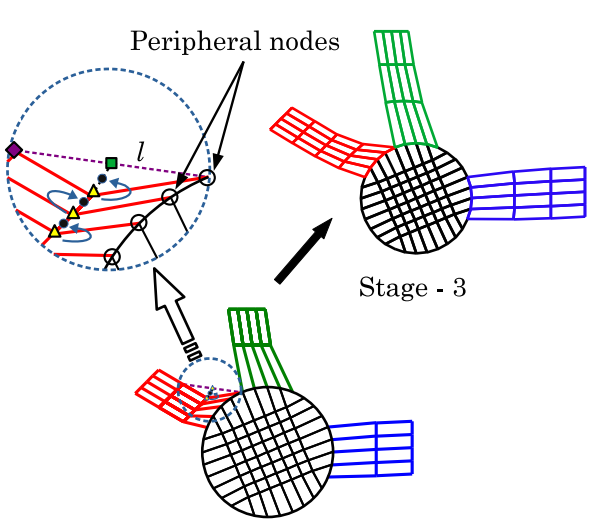}}
			\hfill
			\subcaptionbox
			{Repositioning of nodes of circular junctions for better overall appearance\label{fig:Junction_stage_4}}
			{\includegraphics[scale=1]{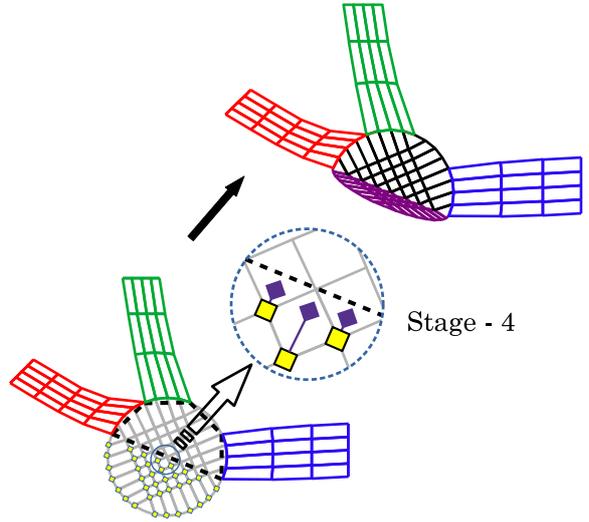}}
			
			\caption{Steps involved in formulating well connected circular junctions. (\subref{fig:Junction_stage_1}) Junction after deleting the overlapping elements of members. (\subref{fig:Junction_stage_2}) Discretized junction connected to the quad node elements of members.  (\subref{fig:Junction_stage_3}) Shifting of node to mid point (\textit{black} point) of the line joining itself (\textit{yellow} triangle) with the midpoint (\textit{green} square)  of another line (dashed \textit{purple} line) formed by joining the adjacent nodes of the quadrilateral elements along the length of the respective member. (\subref{fig:Junction_stage_4}) Re-shaping the circular junctions.}
			\label{fig:mesh}
		\end{figure}
		\subsubsection{Connecting members to junctions}
\begin{figure}[h!]
		 		
		 	\begin{subfigure}{0.49\textwidth}
		 		\centering	
		 		\includegraphics[scale=1]{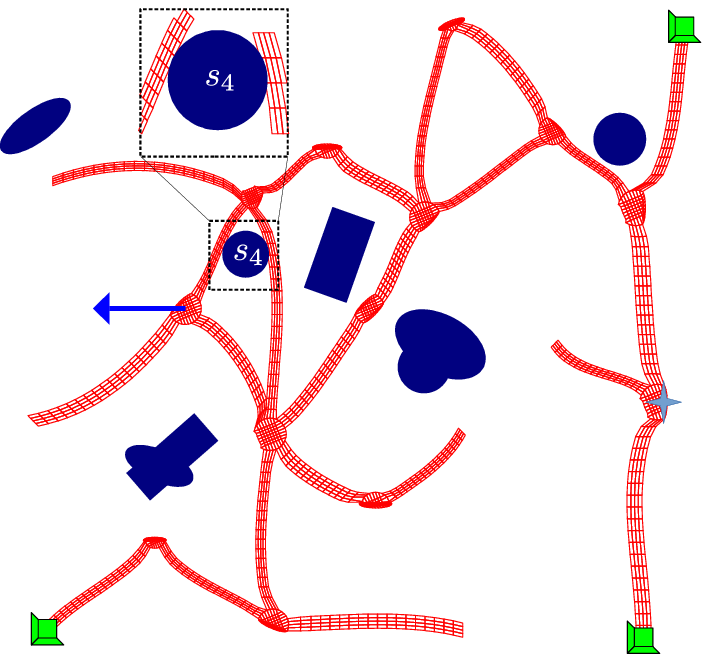}
		 		\caption{ }
		 		\label{fig:Final_Mesh1}	
		 	\end{subfigure}\hfill
		 	\begin{subfigure}{0.49\textwidth}
		 		\centering	
		 		\includegraphics[scale=1]{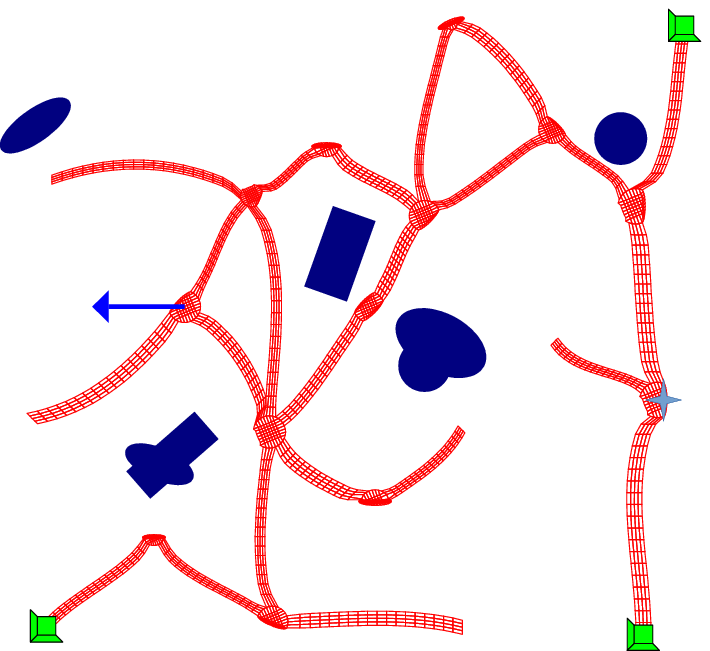}
		 		\caption{ }
		 		\label{fig:Final_Mesh2}	
		 	\end{subfigure}		 
		 	\caption{(\subref{fig:Final_Mesh1}) Mesh shows discretization of frame members with four node quadrilateral elements and reshaped junctions.  External surface ($ S_4 $) intersects with the quadrilateral elements of members, and is ignored in this design. (\subref{fig:Final_Mesh2}) Final mesh with valid external surfaces.}
		 	\label{fig:Final_Mesh}
		 \end{figure}

		The process of connecting quadrilateral elements to circular junctions involves \textit{four} stages. In stage-1, at each junction, all elements whose centroids lie within a circle of radius $R_{ju}$ are deleted as shown in Fig. \ref{fig:Junction_stage_1}. It is important to delete the entire set of elements (\textit{w-set}) even if the centroid of one element belonging to the set falls inside the circular junction. Discretization of circular junction and repositioning of nodes of quadrilateral elements of the design members to ensure connectivity with the respective segment of the circular junction is depicted in Fig. \ref{fig:Junction_stage_2}. One observes that number of elements of all junctions remains same and depends on the number of quadrilateral elements along the width of any design member ($n_{ew} $).  To discretize junctions, every circular junction is mapped to a square with side discretized with $ 2n_{ew} $ elements. Thus, the number of quadrilateral elements of each junction becomes $ 4\times n_{ew}^2 $. Connecting nodes of  quadrilateral elements to the respective peripheral node of the circular junction leads to distortion of the elements (Fig. \ref{fig:Junction_stage_2}). To reduce this undesirable effect, nodes (\textit{yellow} triangles) of quadrilateral elements of the design members that are adjacent to the peripheral nodes (black circles) are re-positioned as shown in Fig. \ref{fig:Junction_stage_3}. A line ($ l $) connecting a peripheral node with the last node (\textit{purple} diamond) of its adjacent element along the length of the member is drawn (\textit{purple} dashed line in Fig. \ref{fig:Junction_stage_3}). Midpoint (black dot) of the line joining the node (\textit{yellow} triangle) adjacent to peripheral node with the midpoint (\textit{green} square) of the newly drawn line ($ l $) is considered as the new nodal position. The methodology proposed here yields satisfactory results though there may be other (and better) methods of avoiding distortion of the quadrilateral elements. Finally, to avoid unintended \textit{bulges} at junctions, nodes of circular junctions not connected to any quadrilateral element of the design members are shifted. An imaginary polygon (represented as \textit{black} dotted line) connecting end nodes of the quadrilateral elements (of design members) that are attached to the peripheral nodes of the circular junction is created as shown in Fig. \ref{fig:Junction_stage_4}. All nodes of the quadrilateral elements whose centroids fall outside this newly created polygon, are moved towards the nearest edge of the polygon based on a compression parameter, considered as 0.75 herein.  In Fig. \ref{fig:Junction_stage_4}, \textit{yellow} diamonds and \textit{purple} squares represent nodes of the elements of the junction before and after shifting towards the nearest edge of the polygon. The above procedure is repeated for all junctions to achieve the final mesh by properly connecting members of the candidate design. To avoid repetition of the above procedure of mesh generation, any external surface that intersects with or encloses a quadrilateral element of any member is flagged and deleted by assigning $v_{s_k}^c$ to 0. In addition, all external surfaces intersecting with the quadrilateral element of any junction are also deleted from the design. Final mesh along with valid members and external surfaces for the candidate design in Fig. \ref{fig:Candidate_solution} is shown in Fig. \ref{fig:Final_Mesh}. External surfaces are also discretized (discretization not shown) using quadrilateral elements.	
 	 
	 	 \subsubsection{Identifying contact surfaces}
\begin{figure}[h!]
		 	\centering	
		 	\includegraphics[scale=1.2]{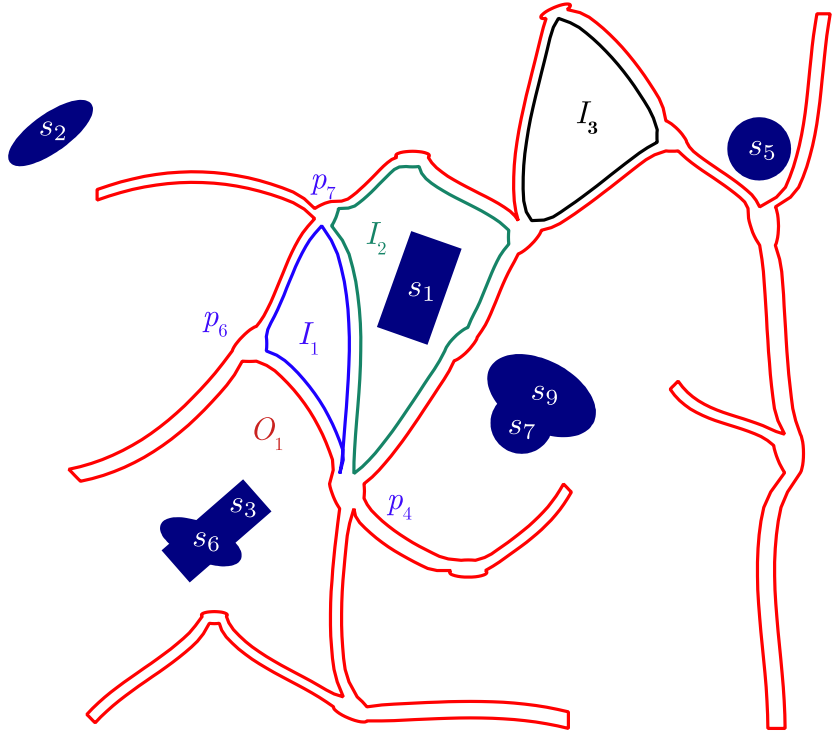} 	
			 	\caption{ Depiction of the CCM continuum as loops. $I_1, I_2 \text{ and } I_3$ represent inner loops. $O_1$ represents the outer loop. $S_1, S_2, S_{3}, S_{5}, S_6,   S_{7} \text{ and }   S_{9}$ represent the external contact surfaces present in the candidate design. Contact interactions between $O_1 \text{ and } S_2, O_1 \text{ and } S_5, O_1 \text{ and } S_{7\cup 9}, O_1 \text{ and } S_{3\cup6}, I_2 \text{ and } S_1$ are specified along with self contact interactions ($O_1$ with $O_1$, $I_1$ with $I_1$, $I_2$ with $I_2$ and $I_3$ with $I_3$) of loops.}
		 	\label{fig:Loops_3}
		\end{figure}

	 	 	It is essential to identify potential contact surfaces, in an efficient way, for large deformation contact analysis. We model each contact surface as a closed loop. A continuum can be approximated by an annular area enclosed within many loops. For example, the mesh in Fig. \ref{fig:Final_Mesh2} is represented as a set of loops (Fig. \ref{fig:Loops_3}) by extracting information from all surface nodes using the proposed algorithm explained below. 

	\subsubsection{Identifying closed loops}\label{Identifying_Loops}
	
	\begin{figure}[h!]
		\centering
		\includegraphics[scale=1]{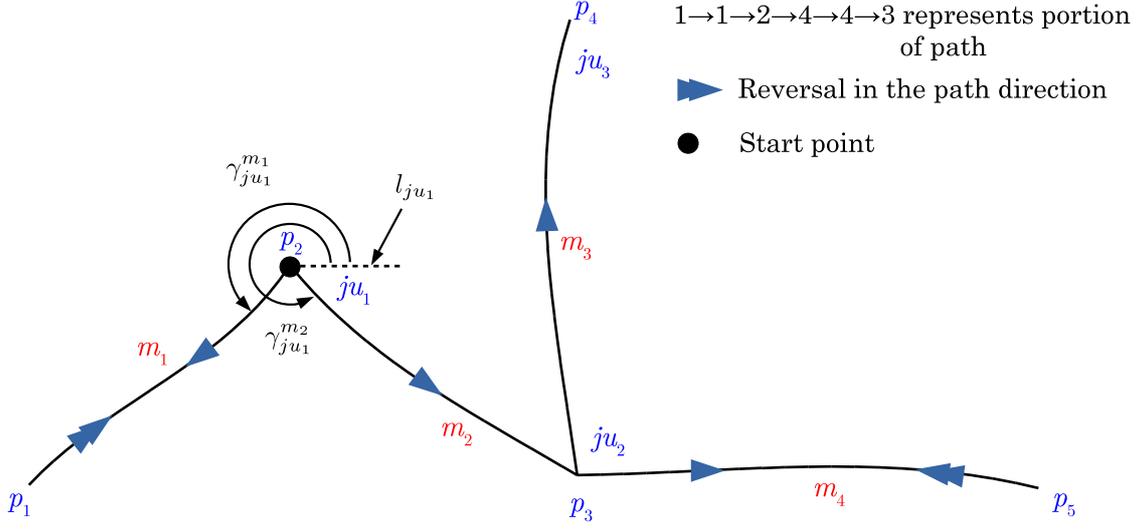} 
		\caption{ Process of identifying members that belong to a path.}
		\label{fig:Loops_0}	
	\end{figure}
	
	\begin{figure}[h!]
		\centering
		\includegraphics[scale=1]{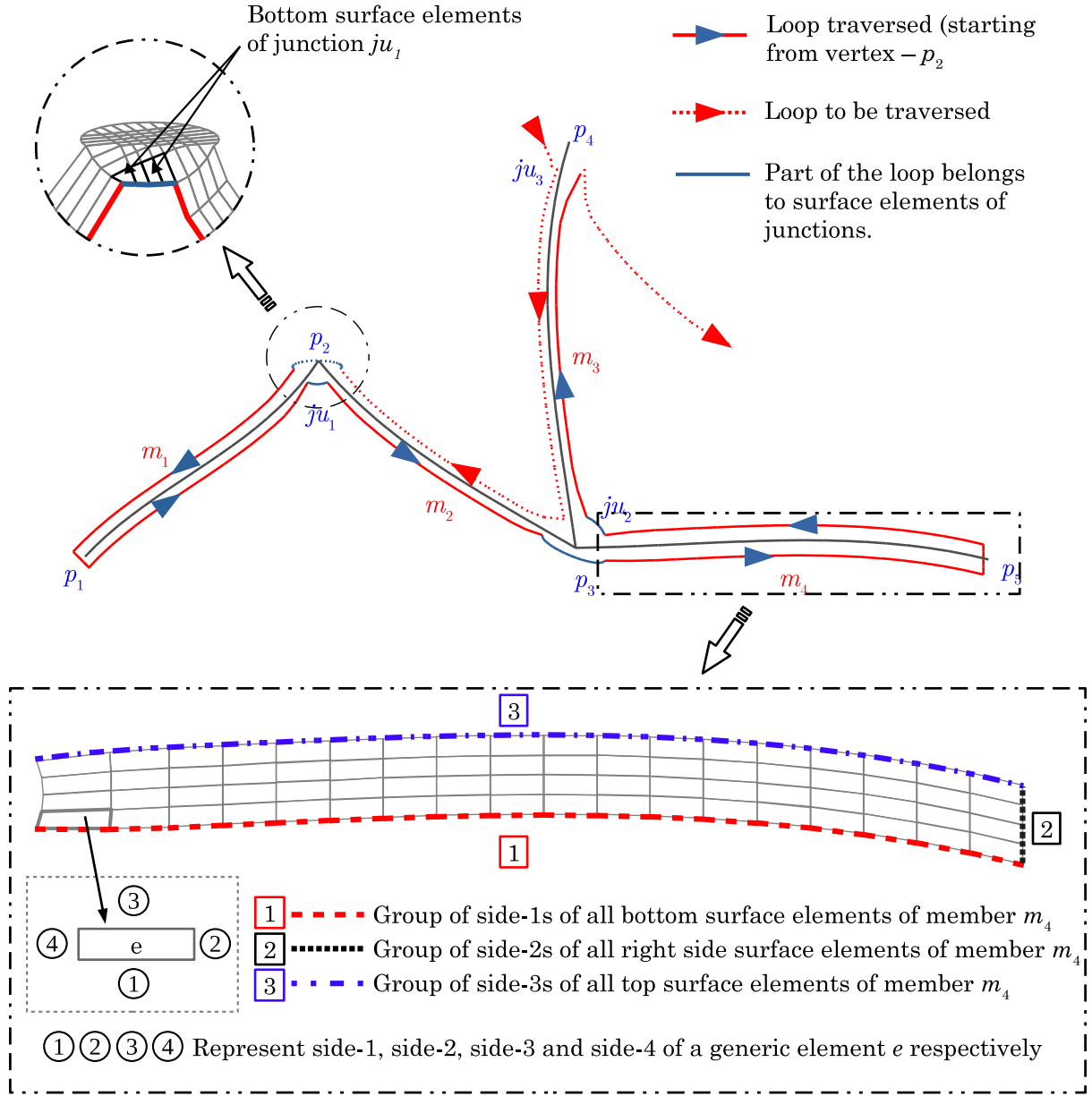} 
		\caption{ Process of identifying surface elements of loops. \textit{Red} continuous line represents portion of the loop traversed while \textit{red} dotted line represents the remaining portion to be traced}
		\label{fig:Loops_1}	
	\end{figure}
	
	\begin{figure}[h!]
		\centering
		\includegraphics[width=0.9\textwidth]{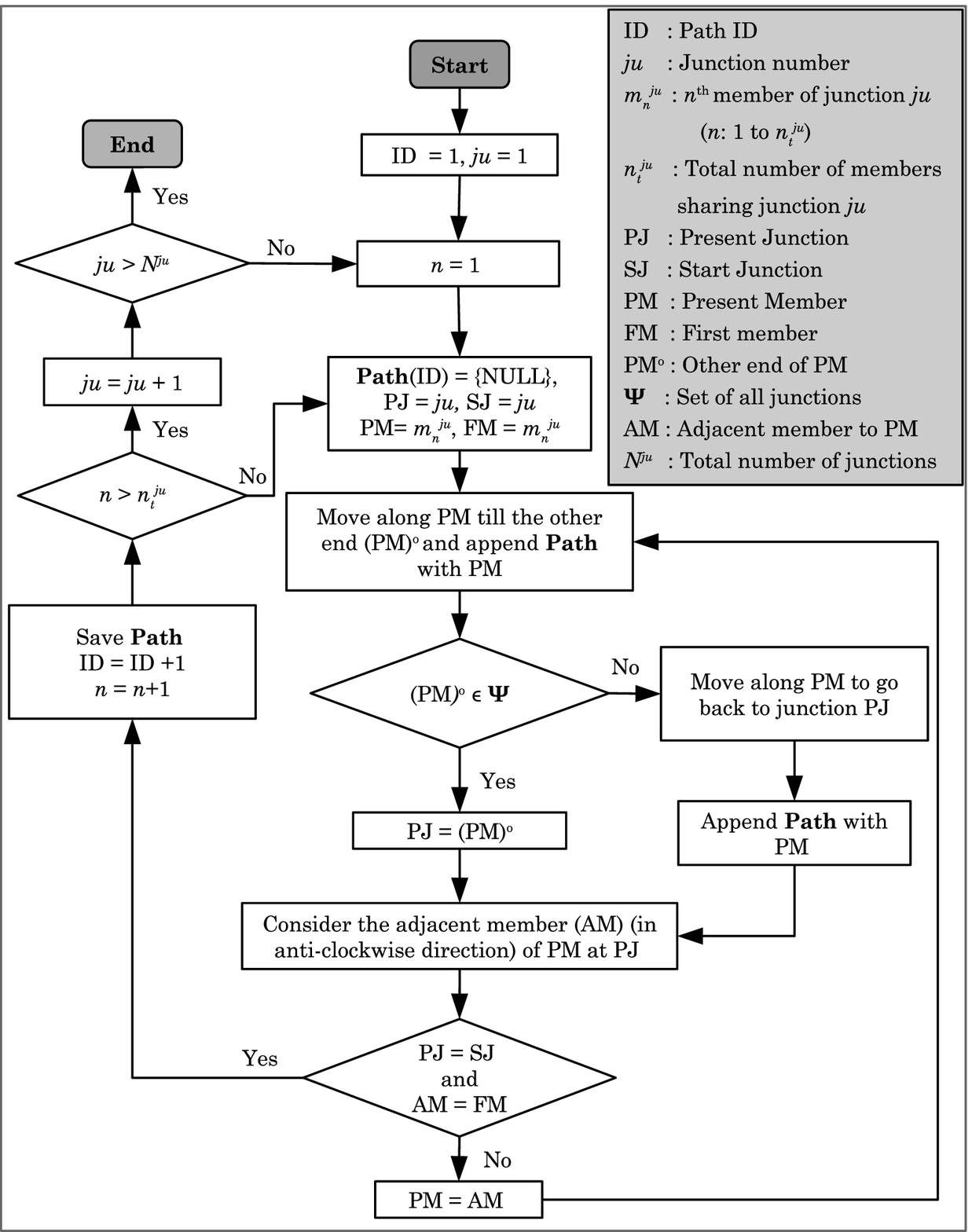} 
		\caption{ Flow chart representing steps involved in the process of identifying paths.}
		\label{fig:Loops_FlowChart} 
	\end{figure}
	
	Closed loops are identified in two stages. In the first, paths are identified by traversing along the neutral axes of members starting from each junction. Each path represents group of members that belong to a loop. In the second, surface elements of members belonging to the identified paths are captured to form loops. 
	
	\noindent \textbf{Stage-1}: To identify paths, we propose the following methodology (Fig. \ref{fig:Loops_FlowChart}). One dimensional wire-frame candidate solution (Fig. \ref{fig:Candidate_solution_2}) without external contact surfaces is chosen as an example to describe how members are identified to form paths. Steps involved are as follows. 
	\begin{enumerate}[label=Step-{\arabic*}]
		\item 
		We chose a junction and commence traversing along the member that makes minimum angle in anti-clockwise direction, among all members sharing that junction, with respect to a horizontal line (in positive $ x- $direction). Once, we reach the other vertex of the traversed member, we append the path with the index of that member, to signify that the member is a part of the corresponding path. For example, consider a portion of 1D candidate solution (Fig. \ref{fig:Candidate_solution_2}) with members $m_1, m_2 \dots, m_4 $ as shown in Fig. \ref{fig:Loops_0}. The search algorithm commences from junction $ ju_1 $ ($ p_2 $) of the wire-frame mesh. From $ ju_1 $, we move along the member $m_1$, since, $ \gamma_{ju_1}^{m_1} < \gamma_{ju_1}^{m_2}  $, where, $ \gamma_{ju_1}^{m_j}$ is the end-slope of member-$ m_j$ with respect to $ l_{ju_1} $ (horizontal line passing through junction ${ju_1} $).  $ m_j, j = 1 \text{ and } 2,$ represent the members sharing ${ju_1} $. After reaching vertex $p_1$ of member $m_1$, we append the path with the index of member $m_1$ i.e., 1.
		
		\item Once at the other end, we check whether it belongs to set of all junctions ($ \Psi $). 
		\begin{enumerate}
			\item If yes, we consider the succeeding member at that vertex (junction), the end-slope of which makes the least angle with respect to that of a traversed member when measured in anti-clockwise direction. Once the succeeding member is identified, we move along it and append the path with its index. 
			\item If the other end is a free end, we move back along the traversed member itself to reach the junction and append its index to path.
		\end{enumerate}
	\end{enumerate}
	Stopping criteria: We repeat Step-2 till we reach the initial junction we started with and the succeeding member is first member of the identified path. Once we satisfy this, we have complete information about one complete path or loop.	We repeat above steps by considering other members sharing the junction.
	
	In Fig. \ref{fig:Loops_0}, As vertex $p_1$ is not a junction, we traverse along member $m_1$ back to junction ${ju_1} $ and append path with index 1 as per step-2(b). Thus the traversed path becomes $ 1\to 1 $. Further, as per step-2(a), we traverse along member $m_2$ since it succeeds $m_1$ at ${ju_1} $ and reach junction ${ju_2} $ (vertex $p_3$) after appending the path with index 2 (path: $ 1\to 1 \to 2 $). The traverse process is continued from ${ju_2}$, and it is identified that member $ m_4 $ succeeds $m_2$. We traverse along $ m_4 $ and append path with index 4 (path: $ 1\to 1 \to 2 \to 4$). We continue  until we satisfy the stopping criteria and the path becomes $ 1\to 1 \to 2 \to 4 \cdots 3\to 2 \to 1$. Further, we repeat the above process considering member $ m_2 $ as the starting member to get path $ 2 \to 4 \to 4 \to 3 \cdots 3\to 2 \to 1 \to 1 \to 2$.

	We follow the above process starting from all junctions of the candidate CCM continuum and identify the remaining paths.  If we choose junction $ ju_2 $ and repeat the above procedure of identifying paths, we get a path $ 4 \to 4 \to 3 \cdots 3 \to 2 \to 1 \to 1 \to 2 \to 4$. With the proposed algorithm, some paths do get repeated. For example, in Fig. \ref{fig:Loops_3}, path corresponding to the inner loop $ I_1$ is traced every time we start from any junction located at vertices $ p_4, p_6 \text{ and } p_7 $.  Repeated paths are flagged off and unique paths used to identify closed loops containing surface elements of the members. We proceed to stage-2 to form closed loops after identifying unique paths.
	
	\noindent \textbf{Stage-2}: Once members belonging to a path are identified, relevant sides of surface quadrilateral elements of members that form a loop are grouped. For example, in Fig. \ref{fig:Loops_1}, side-1s (side number encircled) of all bottom surface elements of member-$ m_4 $ form group-1 (group number enclosed in square; highlighted with \textit{red} dashed line), all side-2s of right side surface elements form group-2 (highlighted with black dotted line) and all side-3s of top surface elements form group-3 (highlighted with \textit{blue} dashed-dotted line). All three groups (1, 2 and 3) collectively  become a part of loop $ O_1 $. This procedure is repeated for all members that belong to the path to get the closed loop $ O_1 $.  Similarly, additional loops are formed by considering the groups of surface quadrilateral elements of members belonging to  each unique path. While considering surface elements of members, those belonging to junctions are also added to get a closed loop. For example, at $ ju_1 $, bottom surface elements of the junction are added to  the loop before considering the bottom surface elements of member $ m_2$ (refer inset in top-left of Fig. \ref{fig:Loops_1}).
	
	After identifying all loops, outer and inner loops are separated to specify contact interactions between loops and external surfaces. In Fig. \ref{fig:Loops_3}, $I_1, I_2,\text{ and } I_3 $ represent the inner loops and $O_1 $ represents the outer loop. All external contact surfaces are also modeled as closed loops by grouping appropriate sides of surface quadrilateral elements. Contact interaction between any external body and the continuum is specified based on position of the contact surface. If any internal loop encloses an external contact surface, contact interaction is specified amongst them. Contact interaction between an external contact surface and the outer loop is specified if the external contact surface lies outside the outer loop. Referring to  Fig. \ref{fig:Loops_3}, it is necessary to specify contact interactions between the external surface $ S_1 $ and inner loop $I_2 $. Similarly, contact interactions between external surface $ S_5 $ and outer loop $O_1 $; external surface $ S_2 $ and outer loop $ O_1$ are also specified. External contact surfaces ($ S_3 \text{ and } S_6 , S_7 \text{ and } S_9 $) overlap to form more complex shapes, thus, promising to offer more intricate deformation characteristics of the continuum, as intended. Interactions of respective loops with such complex external surfaces are also specified. For the candidate design shown in Fig. \ref{fig:Candidate_solution_2} we specify contact interactions between loops $O_1 \text{ and } S_2, O_1 \text{ and } S_5, O_1 \text{ and } S_{7\cup 9}, O_1 \text{ and } S_{3\cup6}, I_2 \text{ and } S_1$ are specified along with self contact interactions ($O_1$ with $O_1$, $I_1$ with $I_1$, $I_2$ with $I_2$ and $I_3$ with $I_3$) of loops.
	
\section{Large displacement contact analysis} \label{Large_Deformation_contact}
	\begin{figure}[h!]
		\centering
		\begin{subfigure}[t]{0.24 \textwidth}
			\centering
			\includegraphics[scale=1]{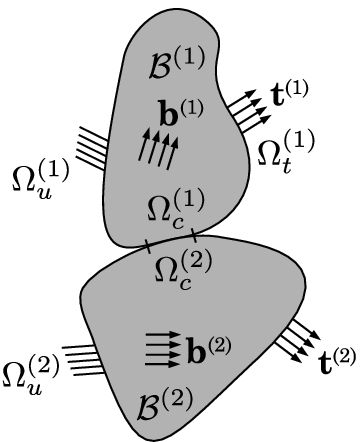}
			\caption{Bodies in deformed configuration}
			\label{fig:contact_bodies}
		\end{subfigure}\hfill
		\begin{subfigure}[t]{0.24 \textwidth}
			\centering
			\includegraphics[scale=1]{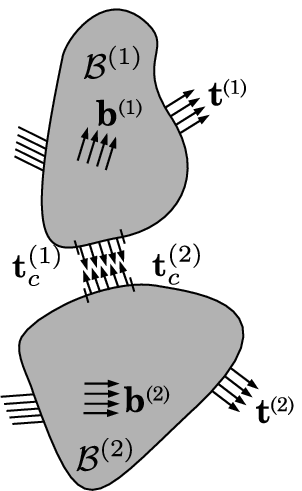}
			\caption{Contact tractions}
			\label{fig:contact_traction}
		\end{subfigure}\hfill
		\begin{subfigure}[t]{0.48 \textwidth}
			\centering
			\includegraphics[scale=1]{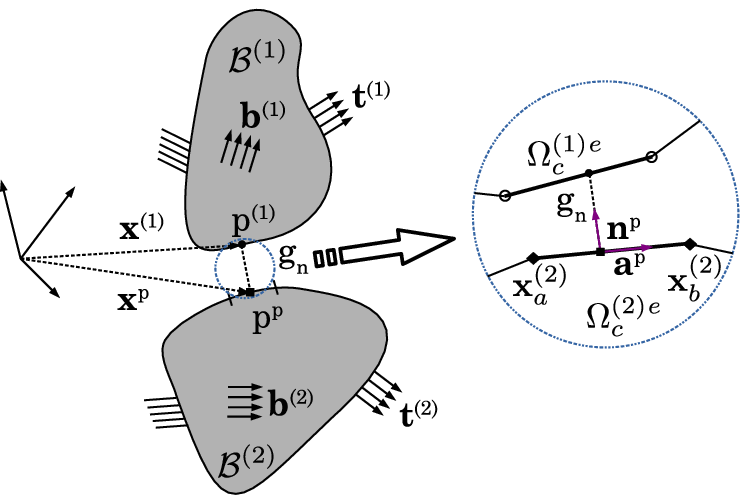}
			\caption{Discretization of contact surfaces as piece-wise linear segments}
			\label{fig:contact_discretization}
		\end{subfigure}
	\caption{Contact modeling}
	\end{figure}
	
\noindent Even though large displacement contact analysis is accomplished using a commercial software, for sake of completeness, the same is briefed below. Consider two bodies $\mathcal{B}^{(\alpha)}$, $ \alpha = 1, 2$ in contact in deformed configuration (Fig. \ref{fig:contact_bodies}). Boundary $ \Omega^{(\alpha)} $ comprises of three segments: $ \Omega^{(\alpha)}_t$ over which surface tractions $ \mathbf{t}^{(\alpha)} $ act;  $ \Omega^{(\alpha)}_u $ with specified displacements $\mathbf{u}^{(\alpha)}$, and $ \Omega^{(\alpha)}_c $ with contact tractions $\mathbf{t}^{(\alpha)}_c$ (Fig. \ref{fig:contact_traction}). Weak form\footnote{For rigid bodies,  $ \nabla\mathbf{u} $ and $ \delta\mathbf{u} $ are both $ \mathbf{0} $, thus, the corresponding contribution to the equilibrium equation becomes zero.} to compute displacement fields $ \mathbf{u}^{(\alpha)} \in \mathcal{R}^{(\alpha)} $ is
	\begin{align}\label{eq:weak_form}
		\begin{split}
		\sum_{\alpha=1}^{2}  
			\underbrace{\int_{\mathcal{B}^{(\alpha)}} \bm{\sigma}^{(\alpha)}:\nabla\left(\delta\mathbf{u}^{(\alpha)}\right)\mathrm{d}v^{(\alpha)}}_{\text{internal virtual work}} 
			- & \sum_{\alpha=1}^{2}  \underbrace{\int_{\Omega^{(\alpha)}_c} \delta\mathbf{u}^{(\alpha)} \cdot \mathbf{t}^{(\alpha)}_c \mathrm{d}a^{(\alpha)}}_{\text{contact virtual work}}\\
			- & \sum_{\alpha=1}^{2}  \underbrace{\left\{\int_{\Omega^{(\alpha)}_t} \delta\mathbf{u}^{(\alpha)} \cdot \mathbf{t}^{(\alpha)} \mathrm{d}a^{(\alpha)} + \int_{\mathcal{B}^{(\alpha)}} \delta\mathbf{u}^{(\alpha)} \cdot \rho^{(\alpha)}\mathbf{b}^{(\alpha)} \mathrm{d}v^{(\alpha)}\right\}}_{\text{external virtual work}}  = 0  \\
			& \qquad \forall \, \delta\mathbf{u}^{(\alpha)} \in \mathcal{S}^{(\alpha)}
		\end{split}
	\end{align}

where $\bm{\sigma}^{(\alpha)}$ is Cauchy stress tensor, $ \nabla\mathbf{u}^{(\alpha)}$ is the gradient of $\mathbf{u}^{(\alpha)}$ with respect to spacial/deformed coordinates, $ \mathbf{b}^{(\alpha)} $ represent body forces per unit volume, $\mathcal{R}^{(\alpha)} $ and $\mathcal{S}^{(\alpha)} $ are permissible sets of deformations and variations respectively, and $ \mathrm{d}a^{(\alpha)} $ and $ \mathrm{d}v^{(\alpha)} $ are elemental areas and volumes in deformed configuration. Elemental displacement field ($ \bm{u}_e$) and its variation ($ \delta\bm{u}_e$) are approximated using finite element shape functions as	
	\begin{align}
		\begin{split}
			\bm{u}_e \approx 		&\sum_{j=1}^{T_n=4}N_j\bm{u}_j = \mathbf{N}\{\mathbf{u}_e\}\\
			\delta\bm{u}_e \approx &\sum_{j=1}^{T_n=4}N_j\bm{v}_j = \mathbf{N}\{\mathbf{v}_e\}\\
		\end{split}
	\end{align}
	
	$ N_j $ represent shape functions for quadrilateral elements with $\mathbf{N} = [N_1\mathbf{I}, N_2\mathbf{I},..., N_{T_n}\mathbf{I}]$, where $ \mathbf{I} $ represents the $ 2 \times 2 $ unit matrix, nodal displacements and variations are represented as $ \bm{u}_j $ and $ \bm{v}_j $ respectively and $\{\mathbf{u}_e\} = [\bm{u}_1^\mathrm{T}, \bm{u}_2^\mathrm{T},...,\bm{u}_{T_n}^\mathrm{T}]^\mathrm{T}$. With these approximations, the following weak form (from Eq. \ref{eq:weak_form}) may be obtained.
	\begin{align}\label{eq:Discretized}
		\bm{v}^\mathrm{T}[\mathbf{f}_{int}+ \mathbf{f}_c-\mathbf{f}_{ext}] = 0 \quad \forall \, \bm{v}\in\mathcal{S}^{(\alpha)}
	\end{align}
	where $ \mathbf{f}_{int}, \mathbf{f}_c \mbox{ and } \mathbf{f}_{ext}$ represent internal, contact and input forces respectively. $ \bm{v} $ comprises variation in nodal displacements $ \bm{u}_j $. Internal elemental forces are $ \mathbf{f}_{int} = \int_{\mathcal{B}^{\alpha}_e} \mathbf{B}^T\bm{\sigma} \mathrm{d}v^{(\alpha)}$, where $ \mathbf{B} $ is strain-displacement matrix. Nonlinear equations $g(\mathbf{u}) \equiv \mathbf{f}_{int}+ \mathbf{f}_c-\mathbf{f}_{ext} = \mathbf{0}$ in terms of the force residual $g(\mathbf{u}) $ derived from Eq. \eqref{eq:Discretized} may be solved either using the Newton Raphson or the Arc Length method \cite{riks1979incremental}. An augmented Lagrangian formulation \cite{Wriggers_2006} is used to compute contact forces ($ \mathbf{f}_c $). Each contact surface is discretized with piece-wise linear segments.
		
	Following is the process to compute contact forces for one body (say, $ \mathcal{B}^{(1)}$)  with respect to the other (say, $ \mathcal{B}^{(2)} $)\footnote{To compute contact forces on $ \mathcal{B}^{(2)}$, the process is reversed.}. Contact tractions are given by 
	\begin{align}
	\label{contact_tractions}
	\mathbf{t}^{(\alpha)}_c =
		\begin{cases}
			\lambda\mathbf{n}^p \qquad & g_n<0\\
			\textbf{0} &g_n \ge 0
		\end{cases}		 
	\end{align}
	where $ \lambda=\lambda_{prev}-\epsilon g_n $ is the Lagrange multiplier modified at every stage from its previous value $\lambda_{prev}$ in the augmented Lagrange setting. $ \epsilon $ is a specified penalty parameter. $ g_n $ is the normal gap given as \begin{align}\label{eq:normal_gap}
			\begin{split}
			g_n = & \mathbf{g}\cdot \mathbf{n}^p\\
			= & \left(\mathbf{x}^{(1)}-\mathbf{x}^p\right)\cdot\mathbf{n}^p
		\end{split}
	\end{align}
	where $ \mathbf{x}^{(1)} $ is the position vector of $ p^{(1)} \in \Omega^{(1)}_c $ (Fig. \ref{fig:contact_discretization}). $ p^p $ is closest projection of $ p^{(1)} $ onto $\Omega^{(2)}_c $, $ \mathbf{x}^p $ is its position vector and $ \mathbf{n}^p $ is normal at $ p^p $. To identify $ \mathbf{x}^p $,   boundary $\Omega^{(2)}_c $ is parameterized using convective coordinates $ \bm{\xi}={\xi^{1},\xi^{2}} $ \cite{wriggers1992coupled}. As we work with 2D elements,   contact surface $\Omega^{(2)}_c $ is approximated as a set of piece-wise linear segments with only $ \xi^p $ as convective coordinate. The tangent ($ \mathbf{a}^p $) at $ p^p $ is computed as $ \mathbf{a}^p = \frac{ \mathbf{x}^{(2)}_b - \mathbf{x}^{(2)}_a}{l^{(2)}_e}$, where, $\mathbf{x}^{(2)}_b, \mathbf{x}^{(2)}_a $ are coordinates of endpoints of segment $ e $ on $ \Omega^{(2)}_c $, and $ l^{(2)}_e $ is the length of segment. $ p^p $ and  $ \xi^p $ are computed by minimizing the distance $ d =||\mathbf{x}^{(1)} - \mathbf{x}^p(\xi^p) || $ as given below. 	
	\begin{align}\label{key}
		\frac{\partial d}{\partial \xi^p} = \left(\mathbf{x}^{(1)}-\mathbf{x}^p(\xi^p)\right). \mathbf{a}^p =0
	\end{align}
	The closest projection point $ \mathbf{x}^p $ on $ \Omega^{(2)}_c $ becomes $ \mathbf{x}^{(2)} $ and the normal gap $ g_n $ is computed using Eq. \eqref{eq:normal_gap}. The $ e^{th} $ segment of the contact surface $ {\Omega^{(2)}_c} $ is approximated using linear shape functions; $ N^{(2)}_a = 0.5(1-\xi^p), N^{(2)}_b = 0.5(1+\xi^p)$. Thus, $ \mathbf{x}^{(2)} = N^{(2)}_a(\xi^p) \mathbf{x}^{(2)}_a + N^{(2)}_b(\xi^p) \mathbf{x}^{(2)}_b $. Elemental contact forces are computed from virtual contact work using contact traction in Eq. \ref{contact_tractions}, and contact stiffness matrix is obtained by differentiating the elemental contact forces with respect to element displacements. 
	
An active set strategy is used to identify and track contact pairs. At each load step, closest projection point is identified and then the normal gap $ g_n $ is computed. For self contact, $ \mathcal{B}^{(1)} $ is the same as $ \mathcal{B}^{(2)} $. Contact forces and stiffness matrices are evaluated if a contact pair is found active, otherwise, they are set to zero if contact is absent or is lost.
	
Implementation of contact analysis delineated above is quite tedious and involved, in particular when designs and contact modes considered are very generic. Many packages, commercial or otherwise, exist that makes themselves amenable and accessible for large deformation analysis with contact modeling. We chose \texttt{ABAQUS\textsuperscript{TM}} to analyze candidate solutions. Procedures highlighted in Sec. \ref{sec:synthesis} for preparation of a candidate design for analysis are all implemented in \texttt{MATLAB\textsuperscript{TM}}. Input data for use in \texttt{ABAQUS\textsuperscript{TM}} pertaining to a candidate CCM design is generated as follows.

	\subsection{ABAQUS\textsuperscript{TM} input file generation} \label{sec:inp_file_gen}
	 \begin{figure}[h!]
		\centering
		\includegraphics[width=\textwidth]{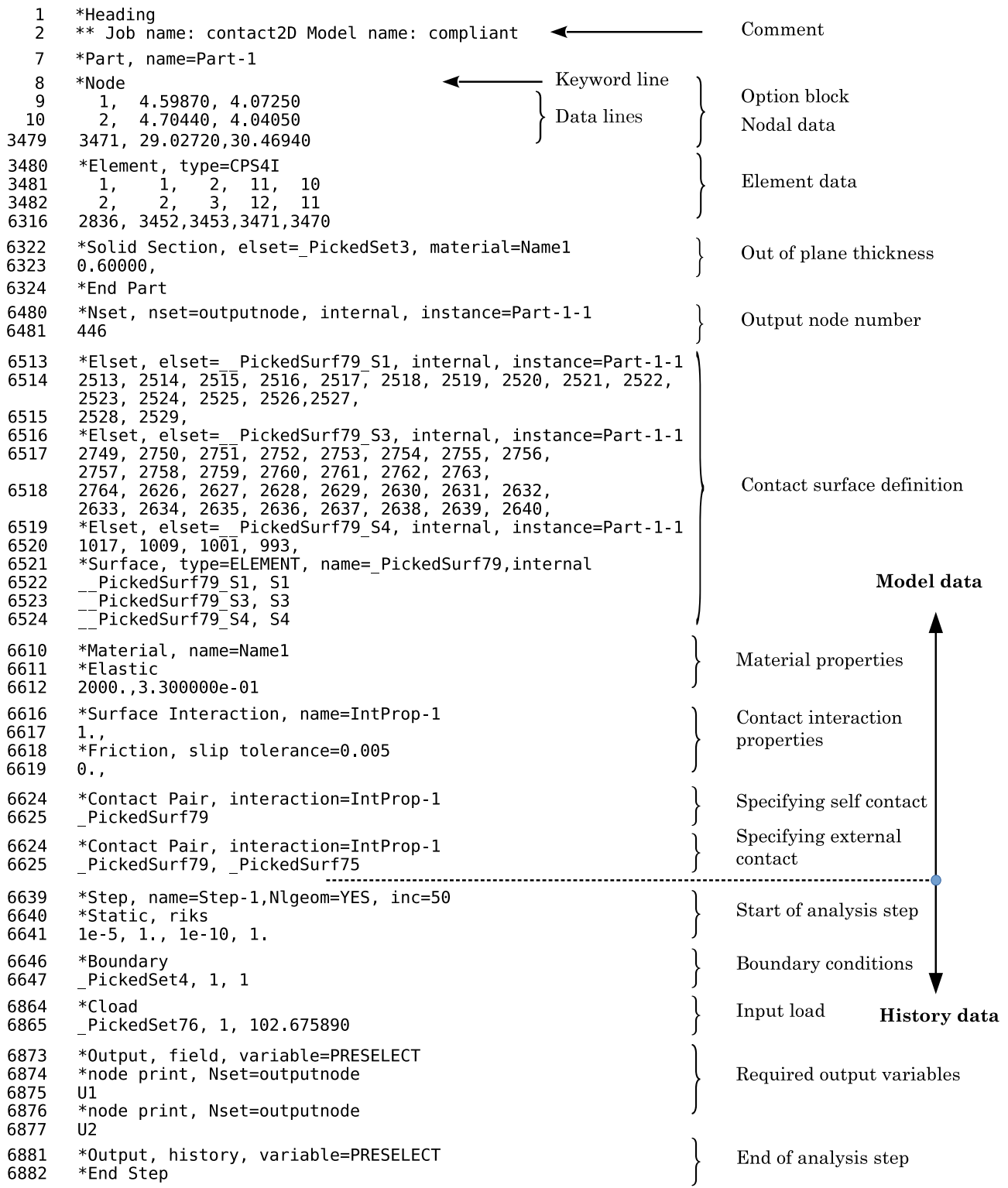} 
		\caption{ Structure of the \texttt{ABAQUS\textsuperscript{TM}} input file. Only select lines are shown for compactness.}
		\label{fig:Input_file} 
	\end{figure}
		 
Mesh information, details of contact surfaces, input force data etc. are prepared using \texttt{MATLAB\textsuperscript{TM}} and supplied to \texttt{ABAQUS\textsuperscript{TM}} solver. \texttt{ABAQUS\textsuperscript{TM}} input file is divided into two main sections. The first part comprises information about the mesh called \textit{model data}. The second part consists of loading conditions and the output data requirements (such as nodal displacements etc) called \textit{history data}. Data of both parts is grouped into \textit{option blocks}. Each option block starts with a \textit{keyword line} and contains several \textit{data lines\footnote{None of these lines can exceed 256 characters.}}. A typical structure of the input file is shown in Fig. \ref{fig:Input_file}. Complete mesh information, i.e., nodal coordinates and element connectivity along with element type, is specified as per the structure of the input file. Further, contact surfaces, i.e., element numbers and corresponding side of the elements are defined, and external and self contact interactions are specified as explained in the previous section. Boundary conditions and input force details along with required output parameters and step information are specified in the history data part as shown in the structure of input file. After preparing the input file as per the standard format, data is analysed by \texttt{ABAQUS\textsuperscript{TM}} standard solver and the output, i.e., the $ x $ and $ y $ displacements of the output port, as sought in output option block of  history data part, is written to a \texttt{.dat} file. Displacement of the output port after each time step is extracted from the \texttt{ABAQUS\textsuperscript{TM}} output(\texttt{.dat}) file. Deformed coordinates of the output port for each time step are compared with the required path, specified by the user,  and objective function is evaluated using Fourier Shape Descriptors as explained in the next section.

\section{Objective and search}
	\label{sec:objective}
	Contact analysis coupled with geometrically nonlinear analysis is performed in \texttt{ABAQUS\textsuperscript{TM}} to obtain the deformation characteristics of a  candidate solution. As we require path generating CCMs to offer repeatability in functionality, we assume that their deformation is in the elastic range. Thus, material nonlinearity is not considered.  Fourier Shape Descriptors (FSDs), that help compare shapes of two curves, are given by  Zahn and Roskies \cite{zahn1972fourier} for curves defined by a set of points forming a closed polygon. Ullah and Kota \cite{ullah1997optimal}, Mankame and Ananthasuresh \cite{mankame2007synthesis}, Rai et al.\cite{rai2007synthesis}, Reddy et al. \cite{Reddy2012systematic} and Kumar et al.\cite{kumar2019computational}  successfully demonstrate the use of FSDs in synthesis of path generating mechanisms. FSDs are obtained by modeling the user specified path and actual output path as closed non-self intersecting curves by chosing an appropriate start point.  
	
Consider $ \alpha_f^d \text{ and } \beta_f^d$ as Fourier coefficients, $l^d \text{ and } \theta^d $ as the length and orientation of the desired path respectively. Similarly, let $ \alpha_f^a \text{ and } \beta_f^a$ represent the Fourier coefficients corresponding to the actual path traced by the candidate solution. Let $l^a \text{ and } \theta^a $ denote the length and orientation of the actual path. The total error objective, $ \mathcal{T}_e $, with necessary weights of individual errors ($ \alpha_e, \beta_e, l_e, \theta_e $), between  actual and desired path is minimized subjected to design variables of both, continuum and the external bodies. Weights corresponding to the errors in shape ($ \varepsilon_\alpha, \varepsilon_\beta $) are kept sufficiently high (100, 100), while, those for errors in length and orientation ($ \varepsilon_l, \varepsilon_\theta $) are chosen low (0.1 and 0) for the examples presented. Reason for choosing zero weight for error in orientation is that a CCM can always be rotated to match initial angles of both the actual and desired paths. The number of Fourier coefficients (N) used for all examples is 100. The objective function $ \mathcal{T}_e $ that is minimized is
	\begin{align*}
		\displaystyle\text{min }\mathcal{T}_e = \varepsilon_\alpha\alpha_e + \varepsilon_\beta\beta_e  &+ \varepsilon_l l_e +\varepsilon_\theta \theta_e 		
	\end{align*}
	\text{where}
	$\displaystyle
	\alpha_e = \sum_{p=1}^N (\alpha_f^d-\alpha_f^a)^2, 
	\beta_e = \sum_{p=1}^N (\beta_f^d-\beta_f^a)^2, 
	l_e		 =  (l^d-l^a)^2,
	\theta_e =  (\theta^d-\theta^a)^2$
			 	
	\noindent over \textbf{v} = 
	$\begin{Bmatrix}
		v_{m_j}^c,v_{s_k}^c ~| \text{ both are discrete variables and take values of either 0 or 1} \\
		v_{s_k}^{sh} ~| \text{ discrete variable and takes values of either 1 or 2 or 3} \\
		v_{m_j}^{T_1}, v_{m_j}^{T_2}, v_{m_j}^w, v^{th}, v_{p_i}^x, v_{p_i}^y, v_{s_k}^x, v_{s_k}^y, v_{s_k}^R, v_{s_k}^{f1}, v_{s_k}^{f2}, v_{s_k}^{\theta}, v_p^F ~| \text{ all continuous}
	\end{Bmatrix}$
			
	\noindent subject to:\vspace{-5mm} 
	\begin{enumerate}
		\item$ \textbf{v}_L \le \textbf{v} \le \textbf{v}_U $ (all continuous variables of \textbf{v} are within the user specified limits).
		\item Equilibrium in large displacement contact analysis.
	\end{enumerate}

A volume constraint is not employed. A random mutation based stochastic Hill Climber search is used to systematically evolve the path generating CCMs. Though computationally costlier compared to a gradient based search, stochastic (function-based) search algorithms have the advantage of penalizing non-convergent solutions. All incomplete intermediate candidate solutions, i.e., candidate designs without either input port, outport or boundary condition, are flagged off by assigning large penalty. Also, both, discrete and continuous variables are handled well by stochastic searches.  {Besides, in the continuum setting (e.g., \cite{kumar2019computational}) and even herein, it is difficult to solve a contact problem with all elements in the void state but physically present, a requirement by the gradient search method. Such void elements will hinder/not permit accurate contact analysis. 

In every iteration, the design vector \textbf{v} is mutated to get new vector $ \textbf{v}^{new} $, according to the mutation probability set by the user, based on a random value generated for each design variable. The total number of variables of the design vector \textbf{v} depends on the intial design domain. For the design domain shown in Fig. \ref{fig:Design_Domain}, the size of design vector \textbf{v} is
	\begin{align*}
		N_v & = 	N_{m} \times 4 + 1 + N_{p} \times 2 + N_s \times 8 + 1\\
	& = 364
	\end{align*}
	where $N_{m}, N_{p}, \text{and } N_s$ represent the total number of members, vertices and external surfaces respectively in the initial design domain.
	
Once the new design vector  ($ \textbf{v}^{new} $) is obtained, the mesh is modified as per Sec. \ref{sec:synthesis},  and the objective function is evaluated. If the new function $ \mathcal{T}_e^{new} $ is less than the earlier function value $\mathcal{T}_e $, $ \textbf{v}^{new} $ is considered and mutated for next iteration; otherwise, $ \textbf{v}$ is mutated further. This is continued until a satisfactory path generating Contact-aided compliant mechanism is obtained.

\section{Example: A three-kink CCM switch}
\label{sec:examples}

\begin{figure}[h!]
	\centering 
		\includegraphics[scale=1.2]{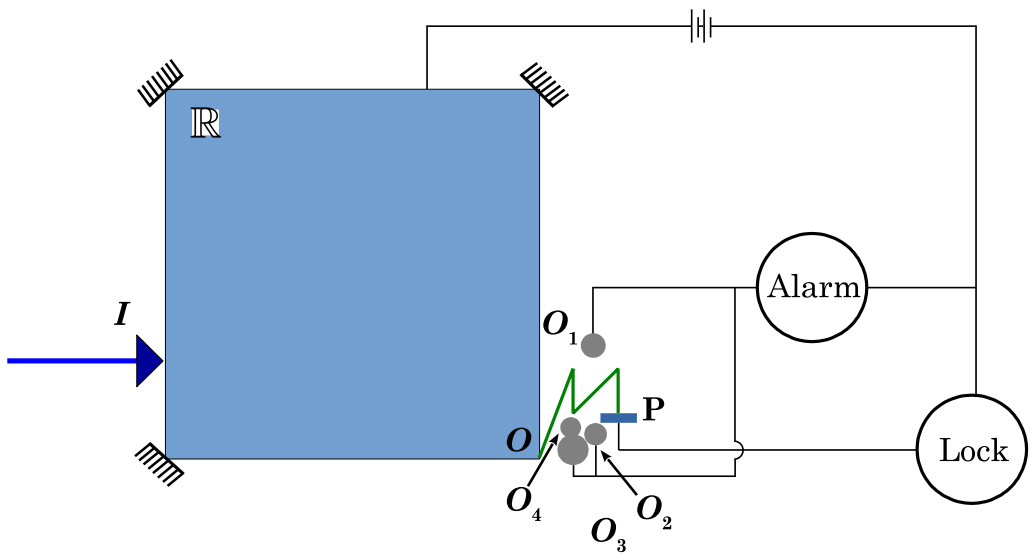}
		\caption{Schematic of the 3-kink mechanical lock. $\mathbb{R}$ represents the region that contains the CCM and key, $ I $ and $ O $ represent input port and output ports respectively. P represents switch pad, $O_1, O_2, \cdots, O_4$ represent obstacles.}
		\label{LockSchematic}
\end{figure}

\begin{figure}[h!]
	\centering\label{fig:Ex_1_1}
	\begin{subfigure}[t]{0.33\textwidth}
		\centering
		\includegraphics[scale=0.65]{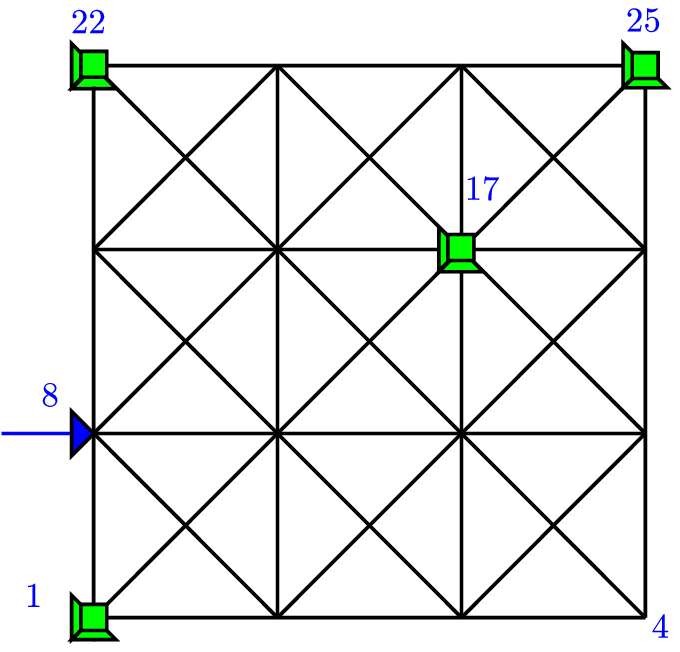}
		\caption{ }
		\label{fig:ex_1_1a}
	\end{subfigure}\hfill
	\begin{subfigure}[t]{0.33\textwidth}
		\centering
		\includegraphics[scale=0.65]{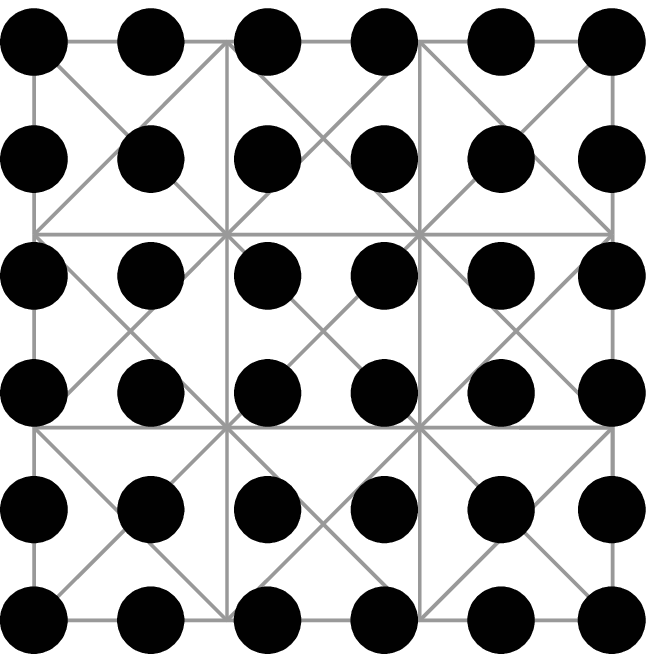}
		\caption{ }
		\label{fig:ex_1b}
	\end{subfigure}\hfill
	\begin{subfigure}[t]{0.33\textwidth}
		\centering
		\includegraphics[scale=0.65]{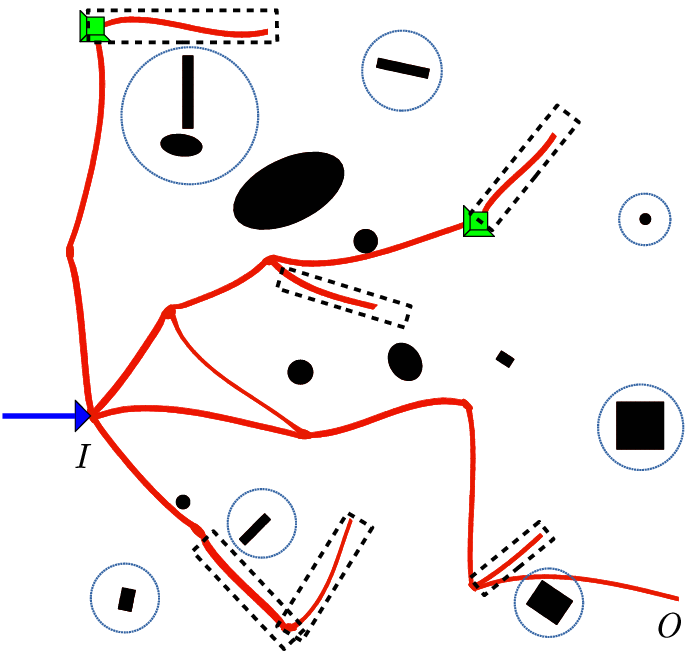}
		\caption{ }
		\label{fig:ex_1_2a}
	\end{subfigure}\\
	
	\begin{minipage}[c]{0.66\linewidth}
		\begin{subfigure}[t]{\textwidth}
			\centering
			\includegraphics[scale=1.24]{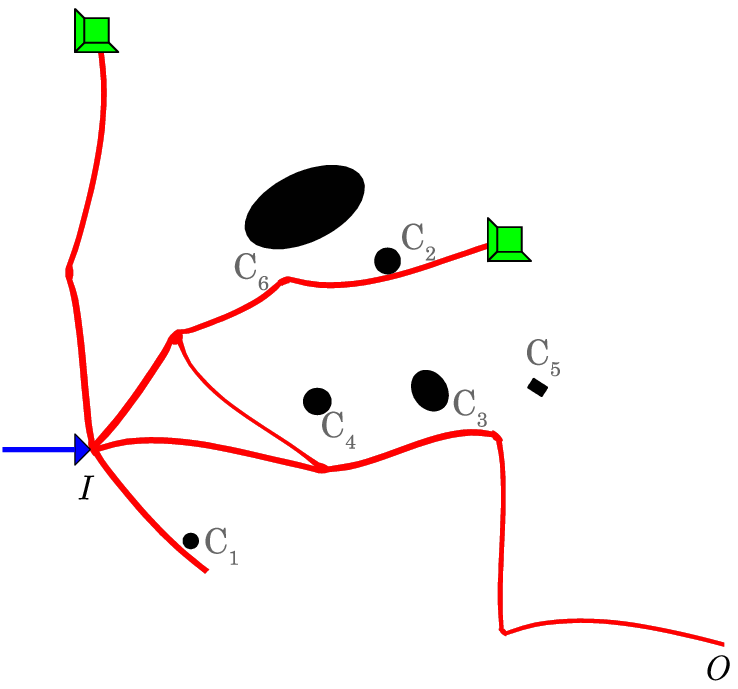}
			\caption{ }
			\label{fig:ex_1_2b}
		\end{subfigure}
	\end{minipage}
	\begin{minipage}[c]{0.33\linewidth}
		\begin{subfigure}[t]{\textwidth}
			\centering
			\includegraphics[scale=0.6]{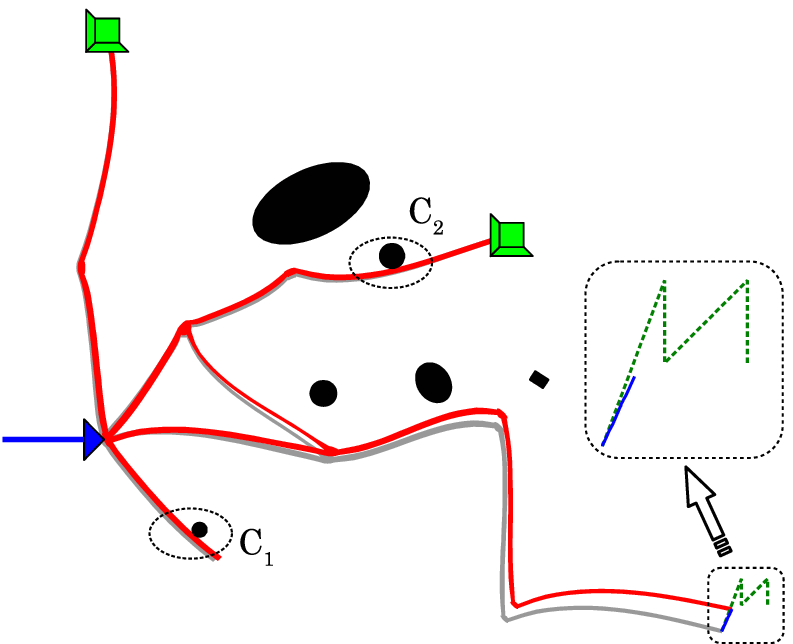}
			\caption{ }
			\label{fig:ex_1_4a}
		\end{subfigure}\\
		\begin{subfigure}[t]{\textwidth}
			\centering
			\includegraphics[scale=0.6]{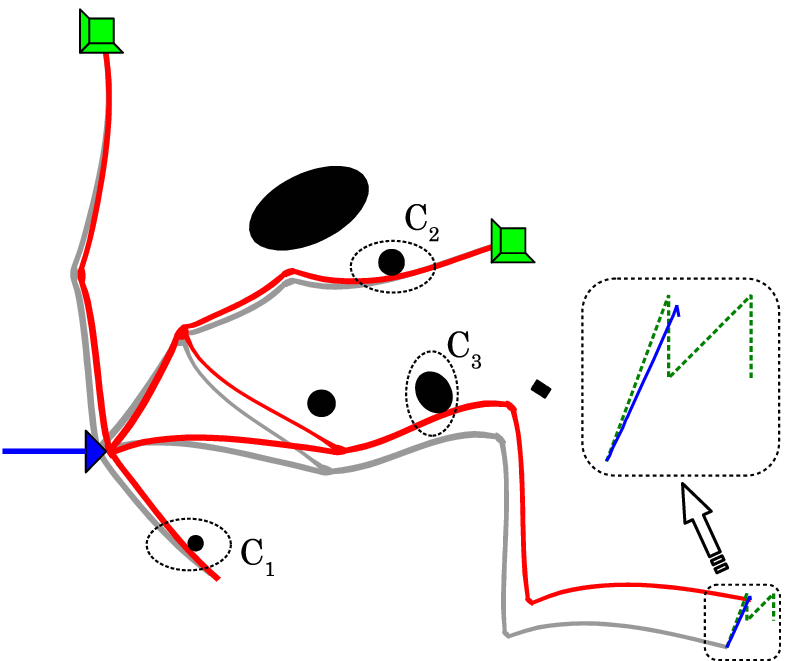}
			\caption{ }
			\label{fig:ex_1_4b}
		\end{subfigure}
	\end{minipage}
	\begin{subfigure}[t]{0.33\textwidth}
		\centering
		\includegraphics[scale=0.65]{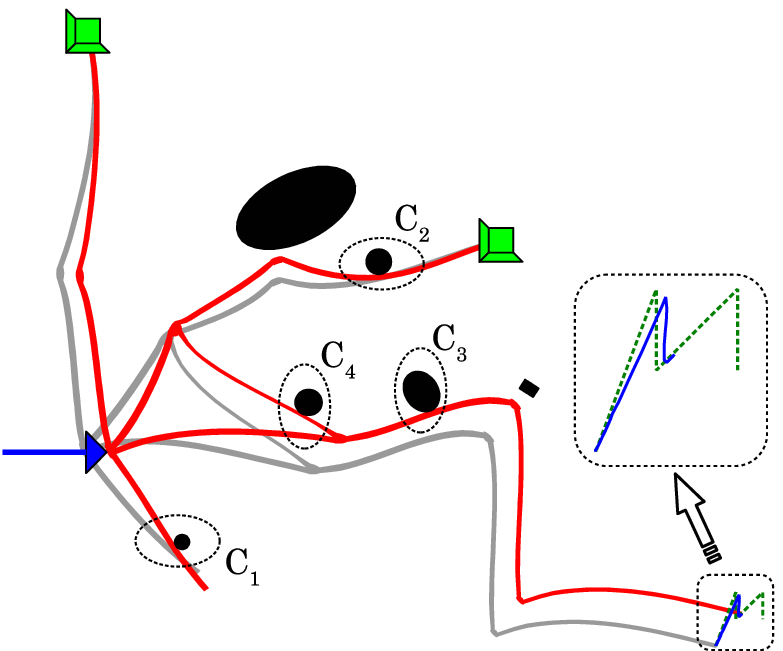}
		\caption{ }
		\label{fig:ex_1_4c}
	\end{subfigure}\hfill
	\begin{subfigure}[t]{0.33\textwidth}
		\centering
		\includegraphics[scale=0.65]{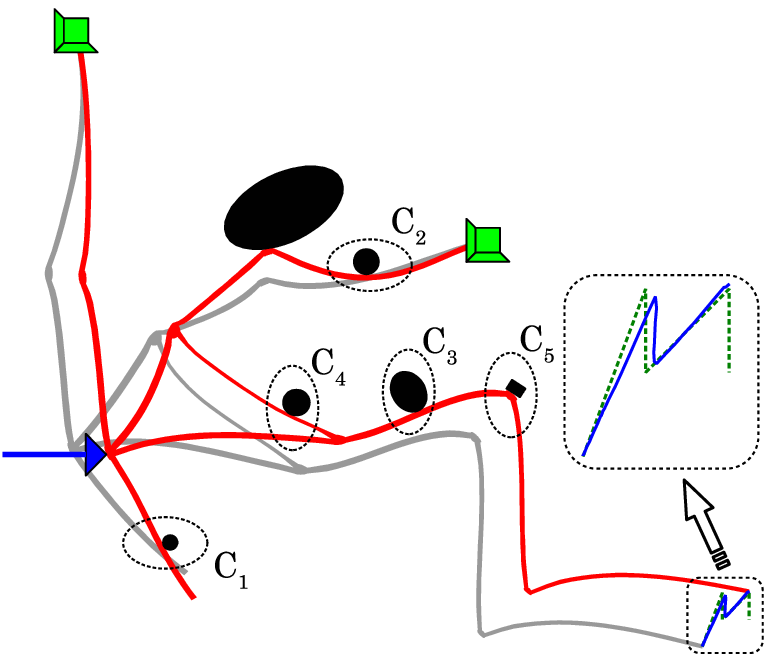}
		\caption{ }
		\label{fig:ex_1_4d}
	\end{subfigure}\hfill
	\begin{subfigure}[t]{0.33\textwidth}
		\centering
		\includegraphics[scale=0.55]{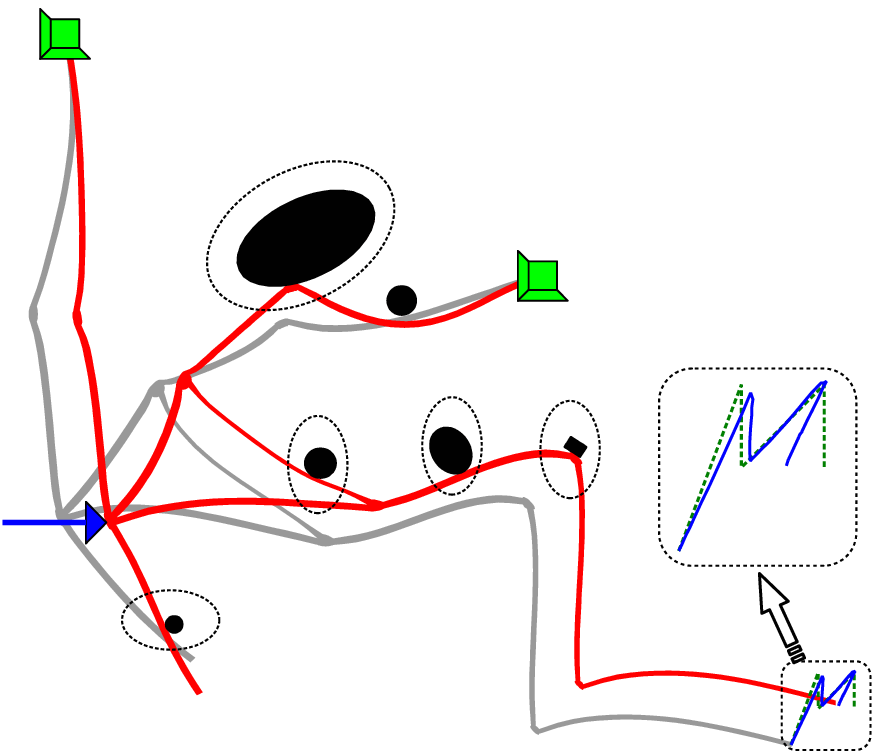}
		\caption{ }
		\label{fig:ex_1_4e}
	\end{subfigure}
	\caption{(a) Initial domain for CCM for the 3-kink path. (b) \textit{Black} circles represent the initial placement of candidate external contact surfaces for the mechanical key. (c) Optimized design of the 3-kink CCM evolved in 1758 iterations along with dangling/non-participating members (enclosed within dashed rectangles) and non-participating external surfaces (encircled). (d) Final design after deleting dangling members and non-participating external surfaces. Members of this CCM are discretized with 3352 quadrilateral elements and 4090 nodes. Continuum in \textit{red} represents deformed and \textit{grey} represents initial/undeformed configurations. Intermediate (e-h) and final deformed configurations (i) of optimized design of CCM. Contact locations are enclosed in dotted ellipses. Desired (dashed \textit{green}) and portion of traced (continuous \textit{blue}) path are shown for each configuration.}
	\label{ThreekinkSwitch}
\end{figure} 

A three-kink electro-mechanical switch is conceived, schematic of which is shown in Fig. \ref{LockSchematic}. The mechanical portion of the switch consists of two parts --- a contact-aided compliant mechanism, and the insulated \textit{key} comprising external surfaces with which the CCM interacts. The CCM and the key are designed simultaneously to lie within the region $\mathbb{R}$, maintained at positive potential. When actuated via an insulated horizontal force at $I$, portion $O$ of the CCM switch is intended to traverse the three-kink path, avoiding obstacles $O_1$, $O_2$ and $O_3$ as shown in the schematic. Prior to its actuation, the mechanical key has to be inserted suitably within $\mathbb{R}$. With a proper mechanical key, the output port of the CCM, after tracing the three-kink path, is expected to make contact with the switch pad for the lock to open. In case the mechanical key is improper, the CCM is expected to deform differently causing the output port to interact with any of the obstacles, and raise an alarm.\\
To seek a CCM and the mechanical key that together generate a  3 kink-path (Fig. \ref{fig:ex_1_3b}), initial domain of size 45 cm $ \times $ 45 cm, with 9 blocks, is considered. Per Fig. \ref{ThreekinkSwitch}a, the design domain comprises 60 members and 25 vertices. Vertices 1, 17, 22 and 25 are fixed, input force is applied at vertex-8 while vertex-4 represents the output port. Magnitude of the input force varies between 0N and 10N. Young's Modulus of continuum is \SI{20}{\newton\per\square\milli\meter} and Poisson's ratio is 0.33. Lower and upper bounds of continuous design variables of members as well as external surfaces are per Tables \ref{Table:Range_v_members} and \ref{Table:Range_v_external}. Optimization parameters used are given in Table \ref{Table:Common_parameters}.
\begin{table}[h]
	\centering
	\caption{Range of design variables for members}
	\label{Table:Range_v_members}
	\begin{tabular}{c P{0.4\textwidth} c c c}	
		\toprule 
		S. No & Design variable & Lower bound & Upper bound & Units\\ 
		\toprule 
		1 & End-slopes $ (v_{m_j}^{T_1}, v_{m_j}^{T_2}) $ & $ - $0.5 & 0.5 & radians \\ 
		2 & In-plane width $(v_{m_j}^w)$& 2 & 6 & mm \\ 
		3 & Out of plane thickness $(v^{th})$& 6& 6 & mm \\ 
		4 & $ x $ and $ y $ coordinates of the bounding box $(v_{p_i}^x, v_{p_i}^y)$ & $ - $20 & 20 & mm \\ 
		\bottomrule 
	\end{tabular}
\end{table}

\begin{table}[h]
	\centering
	\caption{Range of design variables for external surfaces}
	\label{Table:Range_v_external}
	\begin{tabular}{c P{0.4\textwidth} c c c}
		\toprule
		S. No & Design variable & Lower bound & Upper bound & Units\\ 
		\toprule
		1 & $ x $ and $ y $ coordinates of center $(v_{s_k}^x, v_{s_k}^y)$ & 0 & 45 & cm \\ 		
		2 & Radius of bounding circle $ (v^R_{s_k}) $ & 1 & 5 & cm \\ 		
		3 & Size factors $ (v^{f1}_{s_k}, v^{f2}_{s_k}) $ & 0.1 & 1 & ---\\ 		
		4 & Orientation of external surface $ (v^\theta_{s_k}) $ & 0 & $ \pi $ & radians \\ 
		\bottomrule 
	\end{tabular}
\end{table}

\begin{table}[b!]
	\centering
	\caption{Optimization parameters}
	\label{Table:Common_parameters}
	\begin{tabular}{clc}
		\toprule 
		S. No & Parameter                                                                                   & Value \\ \toprule
		1     & Mutation probability                                                                        & 0.08 \\
		2     & Weight of error ($ \alpha_e, \beta_e$) in shape ($ \varepsilon_\alpha, \varepsilon_\beta $) & 100   \SI{}{\per\square\radian}  \\
		3     & Weight of error ($ l_e $) in length ($ \varepsilon_l $)                                     & 0.1  \SI{}{\per\square\milli\meter}\\
		4     & Weight of error ($ \theta_e$) in angle ($ \varepsilon_\theta $)                             & 0   \SI{}{\per\square\radian}     \\
		5     & Number of Fourier coefficients                                                              & 100 \\
		6     & Number of elements along width of member ($n_{ew}$)                                         & 4\\
		7     & Number of elements along length of member ($n_{el}$)                                        & 20\\ \bottomrule
	\end{tabular}
\end{table}

	Fig. \ref{ThreekinkSwitch}b shows the design region along with 36 external contact surfaces, all candidates for the mechanical key. The lower and upper bounds of all continuous variables are chosen to ensure manufacturability. A candidate design (CCM and key) is generated and mutated further as described in section \ref{sec:objective}. The final CCM tracing the desired path, evolved in 1758 iterations, is shown in Fig. \ref{ThreekinkSwitch}c. The magnitude of input force is  \SI{7.35}{\newton}. Dangling members, enclosed within thick dashed black rectangles, and external contact surfaces, enclosed within thin \textit{blue} circles, that do not influence the performance of the switch, are removed. Fig. \ref{ThreekinkSwitch}d shows the final design with the CCM in red and the set of external surfaces comprising the key, in black. Convergence history is depicted in Fig. \ref{fig:ex_1_3a}. Both actual and desired paths are compared in Fig. \ref{fig:ex_1_3b}. Figs. \ref{ThreekinkSwitch}e-i show intermediate deformed configurations of the final design. Contact with external surfaces (key) is observed at six different locations (encircled with dotted ellipse), not all active simultaneously. Sliding contact at  $ \mathrm{C}_1$ and rolling contact at $\mathrm{C}_2 $ are simultaneously active (Fig. \ref{ThreekinkSwitch}e) in the initial load steps. However, no sudden change in output is observed. Further increment in load causes contact at $\mathrm{C}_3 $ (Fig. \ref{ThreekinkSwitch}f) which deviates the direction of output port, thus generating the first kink. Second kink is generated by the sliding contact that becomes active at  $\mathrm{C}_4 $ as shown in Fig. \ref{ThreekinkSwitch}g. Further, successive occurrence of contacts at $\mathrm{C}_5 $ and $\mathrm{C}_6 $ cause the final kink. However, in the mean time contact at $\mathrm{C}_2 $ becomes inactive.

\begin{figure}[h!]
		\centering\label{fig:Ex_1_3}
	\begin{subfigure}[t]{0.48\textwidth}
		\centering
		\includegraphics[scale=1]{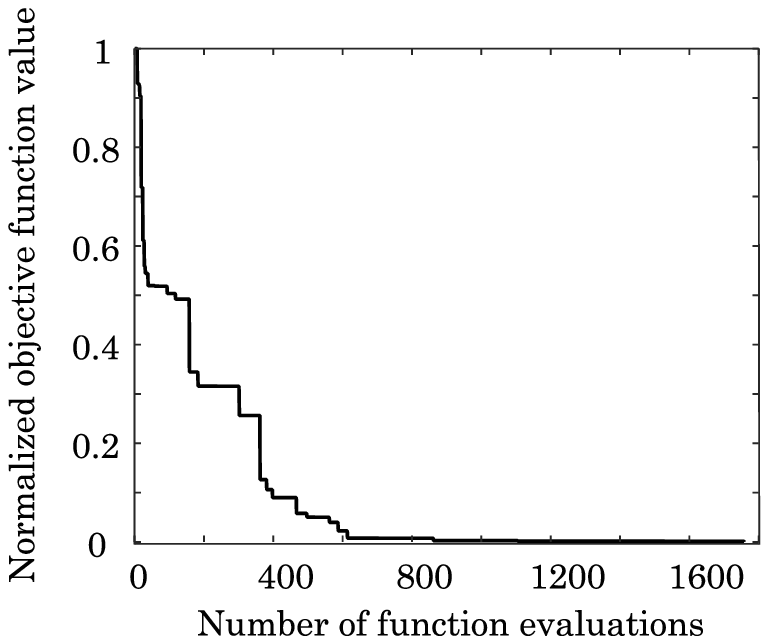}
		\caption{}
		\label{fig:ex_1_3a}
	\end{subfigure}\hfill
	\begin{subfigure}[t]{0.48\textwidth}
		\centering
		\includegraphics[scale=1]{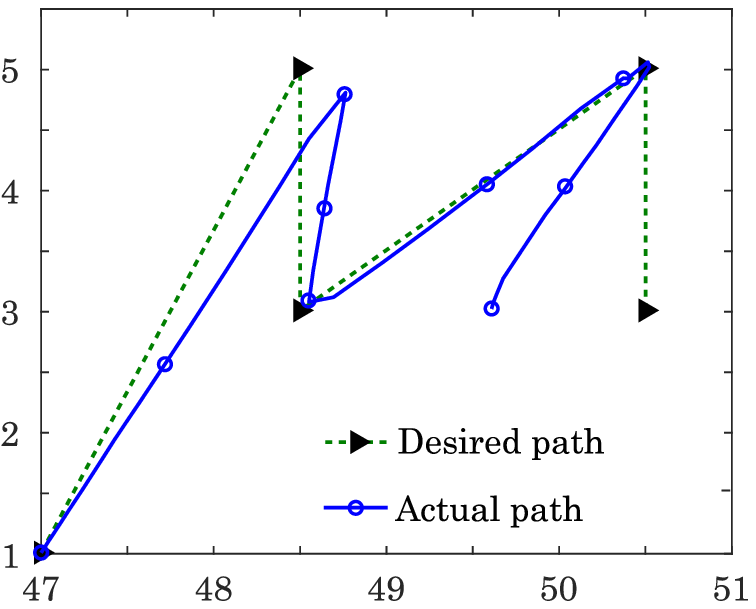}
		\caption{}
		\label{fig:ex_1_3b}
	\end{subfigure}
\caption{(\subref{fig:ex_1_3a}) Convergence history (\subref{fig:ex_1_3b}) Comparison of desired (dashed \textit{green}) and actual (continuous \textit{blue}) paths pertaining to the CCM switch. }
\end{figure}

\section{Discussion}
\label{sec:Discussion}

\subsection{Functionality of the 3-kink switch}
\begin{figure}[t!]
	\centering
	\begin{subfigure}[t]{0.32\textwidth}
		\centering
		\includegraphics[scale=0.84]{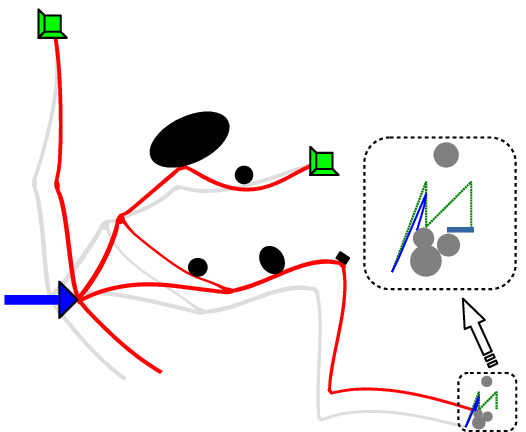}
		\caption{ }
		\label{fig:DWrong_Key_1}
	\end{subfigure}
	\begin{subfigure}[t]{0.32\textwidth}
		\centering
		\includegraphics[scale=0.84]{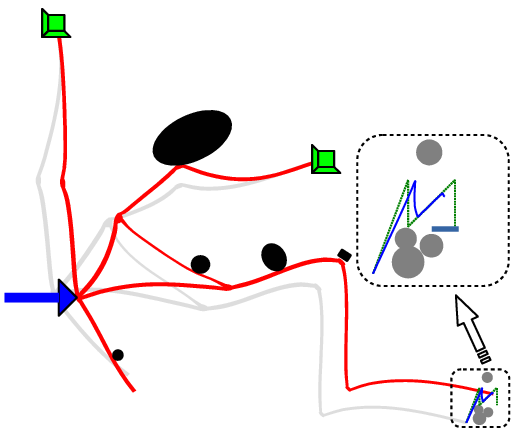}
		\caption{ }
		\label{fig:DWrong_Key_2}
	\end{subfigure}
	\begin{subfigure}[t]{0.32\textwidth}
		\centering
		\includegraphics[scale=0.84]{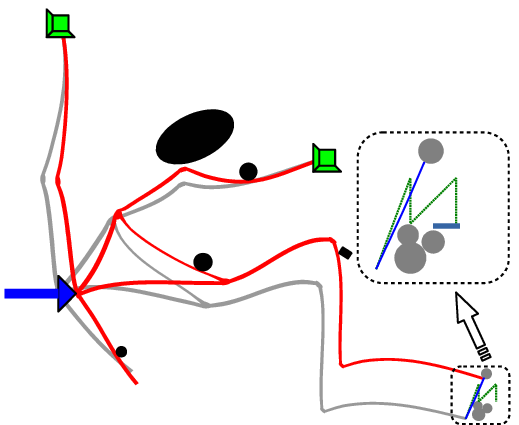}
		\caption{ }
		\label{fig:DWrong_Key_3}
	\end{subfigure}\\
	\begin{subfigure}[t]{0.32\textwidth}
		\centering
		\includegraphics[scale=0.84]{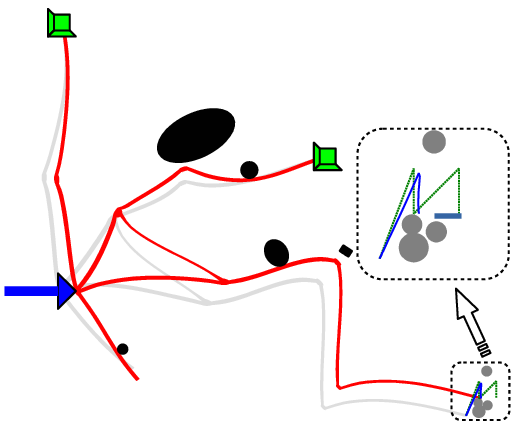}
		\caption{ }
		\label{fig:DWrong_Key_4}
	\end{subfigure}
	\begin{subfigure}[t]{0.32\textwidth}
		\centering
		\includegraphics[scale=0.84]{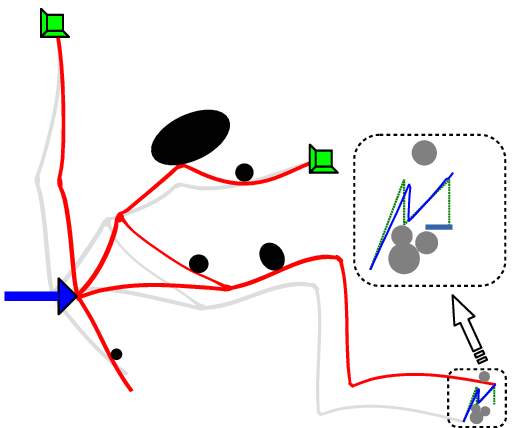}
		\caption{ }
		\label{fig:DWrong_Key_5}
	\end{subfigure}
	\begin{subfigure}[t]{0.32\textwidth}
		\centering
		\includegraphics[scale=0.84]{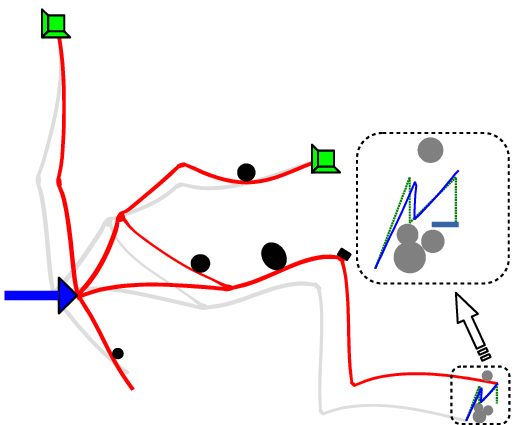}
		\caption{ }
		\label{fig:DWrong_Key_6}
	\end{subfigure}\\
	\begin{subfigure}[t]{0.32\textwidth}
		\centering
		\includegraphics[scale=0.84]{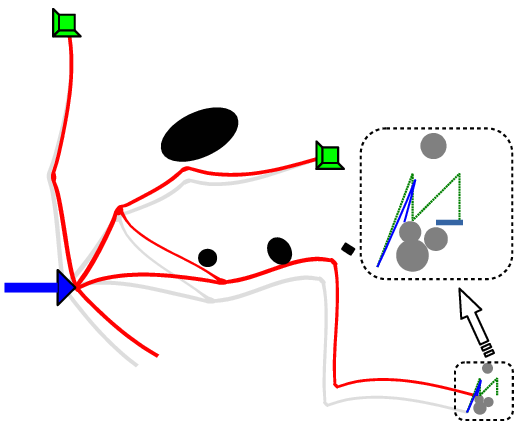}
		\caption{ }
		\label{fig:DWrong_Key_12}
	\end{subfigure}
	\begin{subfigure}[t]{0.32\textwidth}
		\centering
		\includegraphics[scale=0.84]{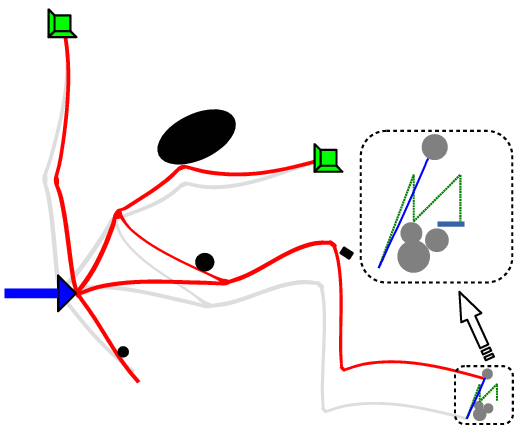}
		\caption{ }
		\label{fig:DWrong_Key_23}
	\end{subfigure}
	\begin{subfigure}[t]{0.32\textwidth}
		\centering
		\includegraphics[scale=0.84]{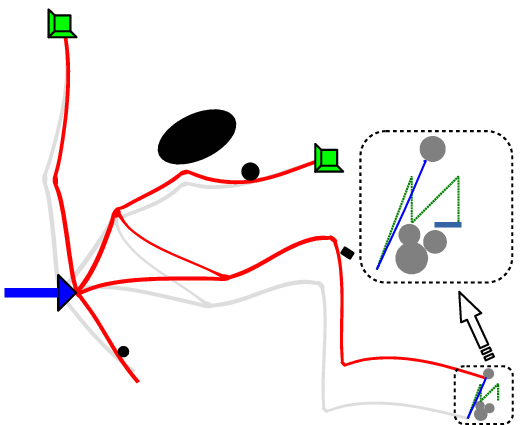}
		\caption{ }
		\label{fig:DWrong_Key_34}
	\end{subfigure}\\
	\begin{subfigure}[t]{0.32\textwidth}
		\centering
		\includegraphics[scale=0.84]{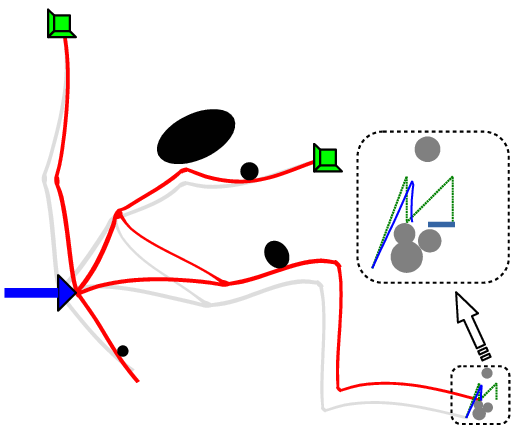}
		\caption{ }
		\label{fig:DWrong_Key_45}
	\end{subfigure}
	\begin{subfigure}[t]{0.32\textwidth}
		\centering
		\includegraphics[scale=0.84]{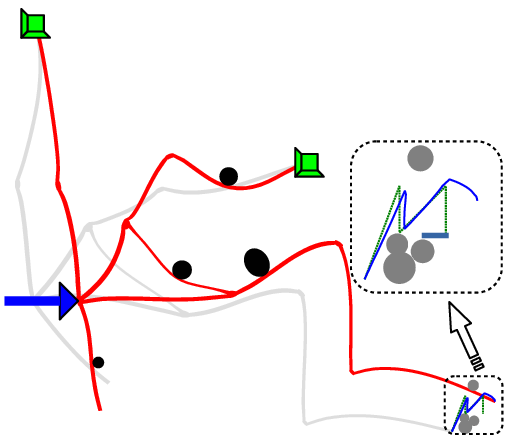}
		\caption{ }
		\label{fig:DWrong_Key_56}
	\end{subfigure}
	\begin{subfigure}[t]{0.32\textwidth}
		\centering
		\includegraphics[scale=0.84]{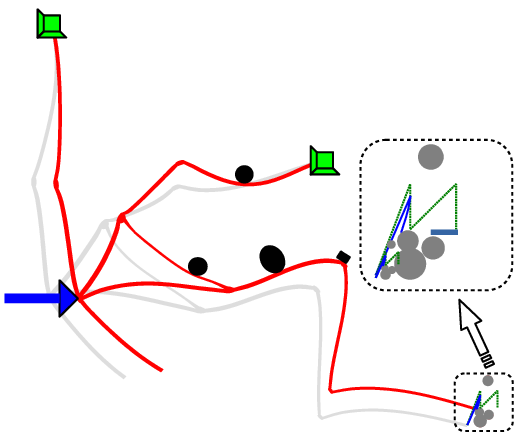}
		\caption{ }
		\label{fig:DWrong_Key_61}
	\end{subfigure}\\
	\begin{subfigure}[t]{0.32\textwidth}
		\centering
		\includegraphics[scale=0.84]{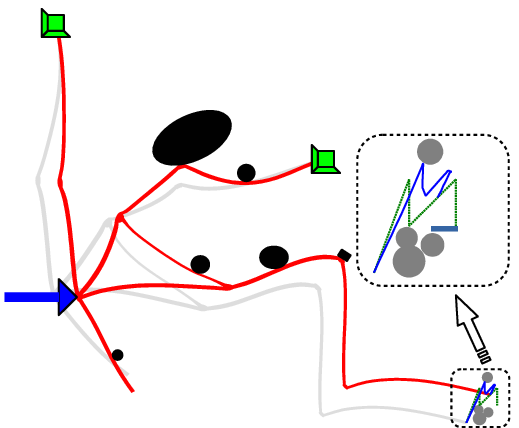}
		\caption{ }
		\label{fig:DWrong_Key_R3}
	\end{subfigure}
	\begin{subfigure}[t]{0.32\textwidth}
		\centering
		\includegraphics[scale=0.84]{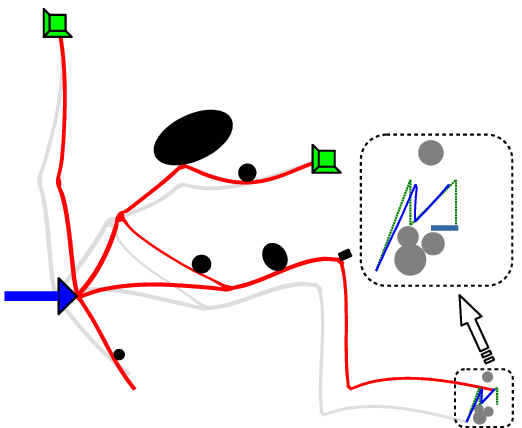}
		\caption{ }
		\label{fig:DWrong_Key_R5}
	\end{subfigure}
	\begin{subfigure}[t]{0.32\textwidth}
		\centering
		\includegraphics[scale=0.84]{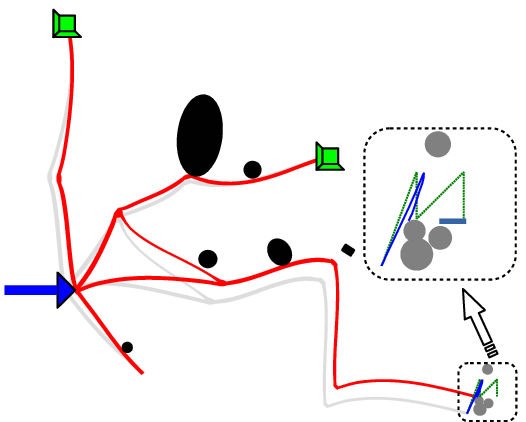}
		\caption{ }
		\label{fig:DWrong_Key_R6}
	\end{subfigure}
	\caption{Wrong key combinations. (\subref{fig:DWrong_Key_1}-\subref{fig:DWrong_Key_6}) represent the absence of an external contact surface one at a time. (\subref{fig:DWrong_Key_12}-\subref{fig:DWrong_Key_61}) represent the wrong key combinations where two external contact surfaces are absent. (\subref{fig:DWrong_Key_R3}-\subref{fig:DWrong_Key_R6}) represent the combinations of wrong key when a non-circular external contact surface is rotated by `one' radian in anti-clockwise direction about its center. Deformed positions of continuum are shown in \textit{red} and undeformed configurations are shown in \textit{light gray}. Obstacles are shown as \textit{gray} circles.} \label{fig:WrongKeys}
\end{figure}

The switch is operated in two stages: (i) the key comprising the external surfaces is inserted, and then (ii) desired input force is applied at the input port of the switch via an insulated servo motor. On insertion of the correct key, i.e., positions, sizes, types and orientations of external contact surfaces are as per the design, the applied force causes the CCM to deform and interact with external contact surfaces. The output port traces the 3-kink path, avoiding the obstacles, and makes contact with the switch pad to open the lock. However, if a wrong key is inserted, the output port does not trace the desired path. Use of a wrong key can cause the continuum to deform in a different manner, which may result in the output port making contact with any obstacle and trigger a security alarm system. Though one can formulate numerous key combinations, we chose 15 candidate wrong key combinations as shown in Fig. \ref{fig:WrongKeys} to demonstrate functionality of the switch. These combinations are achieved by removing either one (Figs. \ref{fig:WrongKeys}\subref{fig:DWrong_Key_1}-\subref{fig:DWrong_Key_6}) or two  (Figs. \ref{fig:WrongKeys}\subref{fig:DWrong_Key_12}-\subref{fig:DWrong_Key_61}) external contact surfaces and/or rotating (Figs. \ref{fig:WrongKeys}\subref{fig:DWrong_Key_R3}-\subref{fig:DWrong_Key_R6}) some non-circular external surfaces by `one' radian in counter-clockwise direction. In all chosen combinations, one observes that the output port does not reach the switch pad, thus keeping the lock intact. However, in three cases (Figs. \ref{fig:WrongKeys}\subref{fig:DWrong_Key_2}, \subref{fig:DWrong_Key_56} and \subref{fig:DWrong_Key_R5}), neither the output port nor any member of the CCM touches any of the obstacles and does not trigger the safety alarm. Nevertheless, the lock does not get open. In all other cases,  deformation of the CCM is such that the alarm system is triggered. It is observed in most cases that either the output port touches the obstacle-1 ($ O_1 $) or obstacle-4 ($ O_4 $) or the member attached to output port touches the obstacle-1 ($ O_1 $) and raises the alarm.

\subsection{Effect of friction}
\label{sec:friction}
The 3-kink switch in Fig. \ref{ThreekinkSwitch}d is designed with zero friction assuming the contact surfaces are very smooth and well polished. In case friction exists, we chose different coefficients, $ \mu = 10^{-3}, 10^{-2}, 0.05$ and $10^{-1}$  at all contact-sites and analyse the final design. Behavior of the 3-kink CCM with respect to different coefficients of friction is depicted in Figs. \ref{fig:Friction}\subref{fig:Friction_mu_0_001}-\subref{fig:Friction_mu_0_1}. One observes that even though the paths are not very different from that desired, the input force required is quite high, and increases significantly with the coefficient of friction. It is suggested that the mechanical key has very smooth external surfaces ($C_1$ to $C_6$ in Fig. \ref{ThreekinkSwitch}d) to maintain low friction coefficients for the output port to reach the switch pad with minimum effort.

\begin{figure}[h!]
	\centering
	\begin{subfigure}[t]{0.24\textwidth}
		\centering
		\includegraphics[scale=0.5]{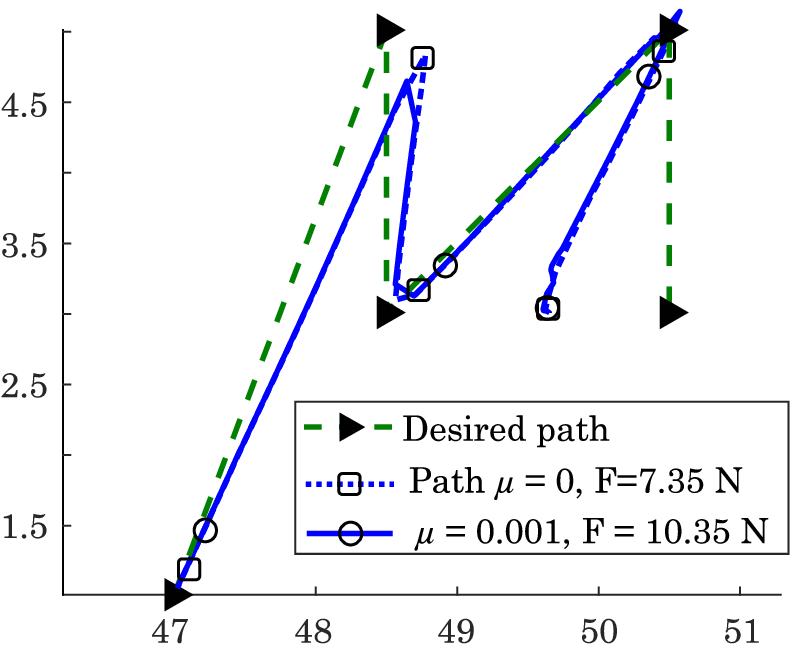}
		\caption{$ \mu = 0.001, F =  10.35 $N}
		\label{fig:Friction_mu_0_001}
	\end{subfigure}\hfill
	\begin{subfigure}[t]{0.24\textwidth}
		\centering
		\includegraphics[scale=0.5]{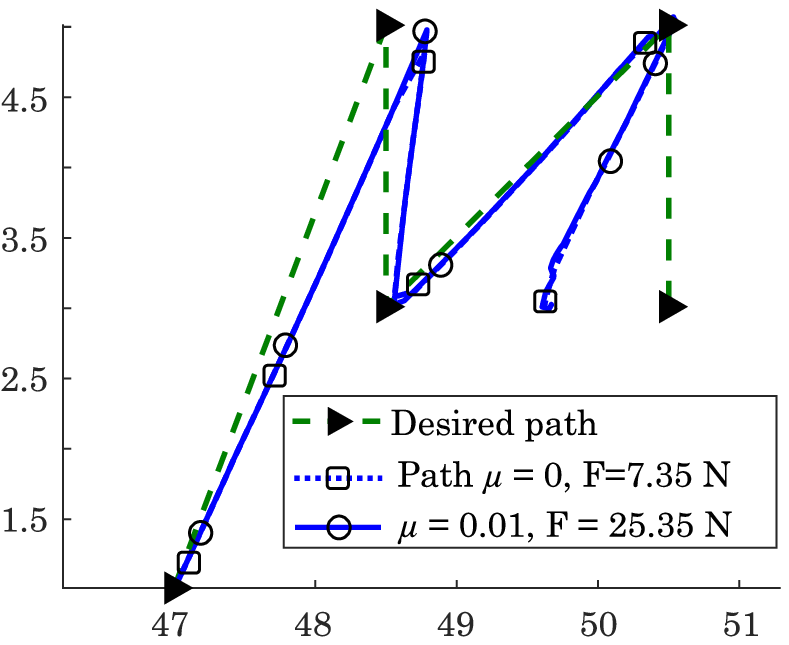}
		\caption{$ \mu = 0.01, F =  25.35 $N}
		\label{fig:Friction_mu_0_01}
	\end{subfigure}\hfill
	\begin{subfigure}[t]{0.24\textwidth}
		\centering
		\includegraphics[scale=0.5]{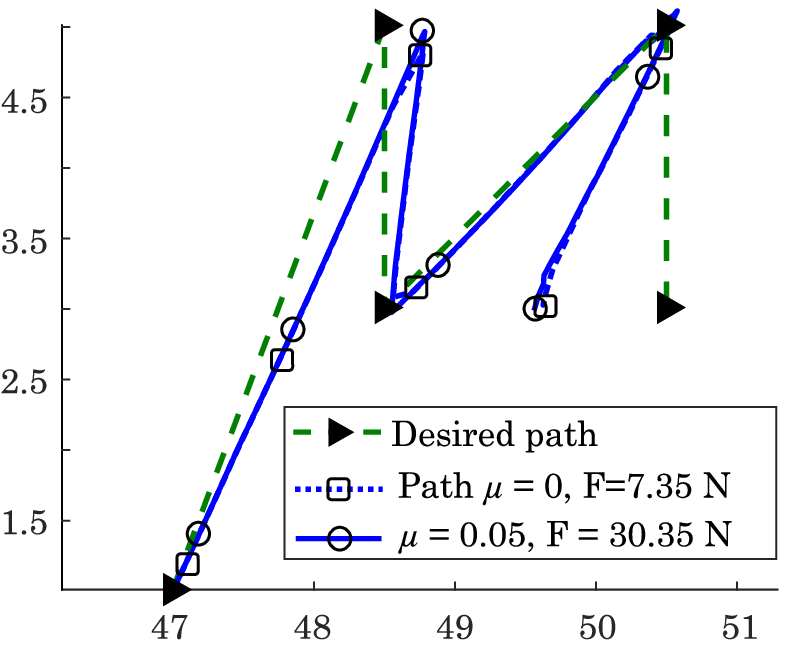}
		\caption{$ \mu = 0.05, F =  30.35 $N}
		\label{fig:Friction_mu_0_05}
	\end{subfigure}\hfill
	\begin{subfigure}[t]{0.24\textwidth}
		\centering
		\includegraphics[scale=0.5]{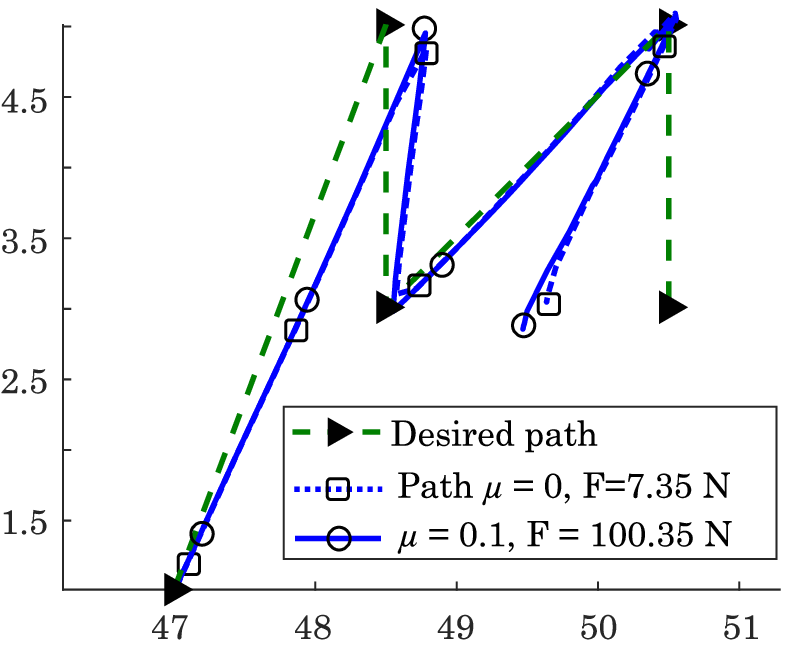}
		\caption{$ \mu = 0.1, F =  100.35 $N}
		\label{fig:Friction_mu_0_1}
	\end{subfigure}
	\begin{subfigure}[t]{0.24\textwidth}
		\centering
		\includegraphics[scale=0.55]{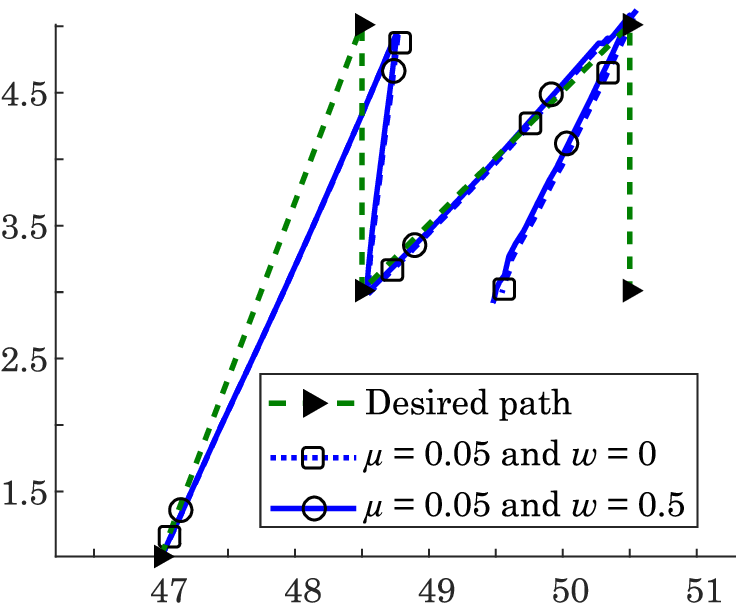}
		\caption{$ w $ = 0.5\%, $ \mu = 0.05 $}
		\label{fig:Wear_mu_0.05_w_0.5}
	\end{subfigure}\hfill
	\begin{subfigure}[t]{0.24\textwidth}
		\centering
		\includegraphics[scale=0.55]{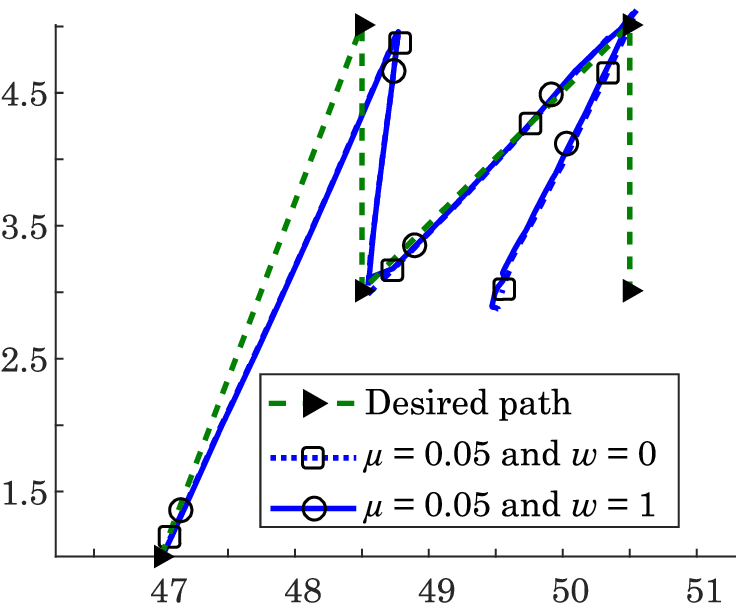}
		\caption{$ w $ = 1.0\%, $ \mu = 0.05 $}	
		\label{fig:Wear_mu_0.05_w_1}
	\end{subfigure}\hfill
	\begin{subfigure}[t]{0.24\textwidth}
		\centering
		\includegraphics[scale=0.55]{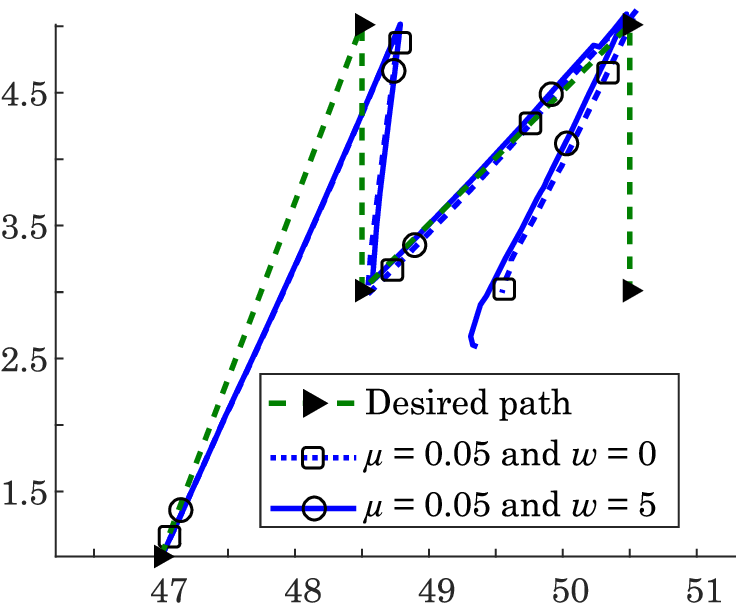}
		\caption{$ w $ = 5.0\%, $ \mu = 0.05 $}
		\label{fig:Wear_mu_0.05_w_5}
	\end{subfigure}\hfill
	\begin{subfigure}[t]{0.24\textwidth}
		\centering
		\includegraphics[scale=0.55]{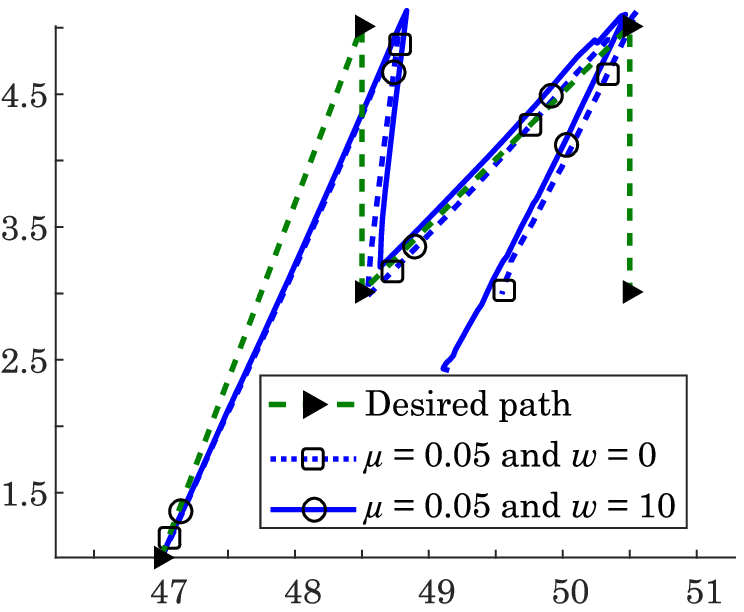}
		\caption{$ w $ = 10\%, $ \mu = 0.05 $}
		\label{fig:Wear_mu_0.05_w_10}
	\end{subfigure}
	\begin{subfigure}[t]{0.24\textwidth}
		\centering
		\includegraphics[scale=0.6]{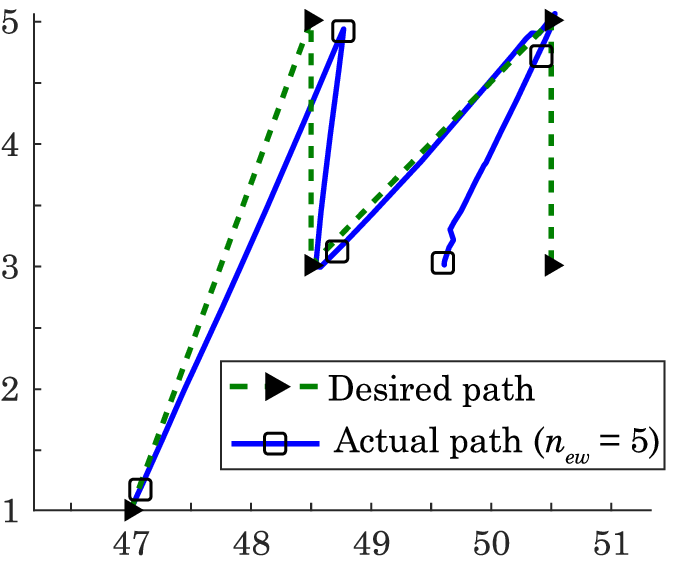}
		\caption{$ n_{ew} = 5 $}
		\label{fig:new_5_nel_20}
	\end{subfigure}\hfill
	\begin{subfigure}[t]{0.24\textwidth}
		\centering
		\includegraphics[scale=0.6]{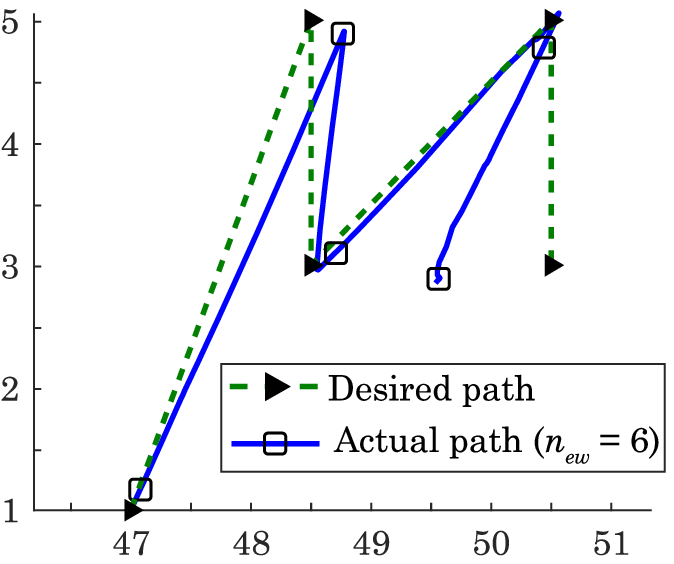}
		\caption{$ n_{ew} = 6 $}		
		\label{fig:new_6_nel_20}
	\end{subfigure}\hfill
	\begin{subfigure}[t]{0.24\textwidth}
		\centering
		\includegraphics[scale=0.6]{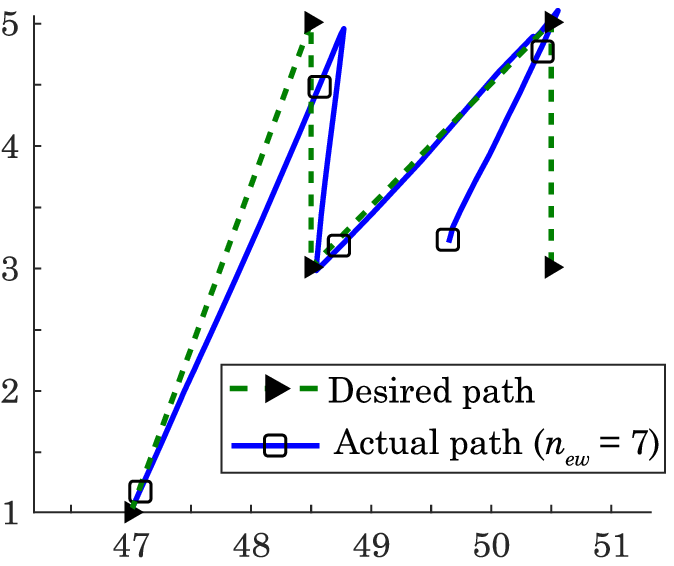}
		\caption{$ n_{ew} = 7 $}
		\label{fig:new_7_nel_20}
	\end{subfigure}\hfill
	\begin{subfigure}[t]{0.24\textwidth}
		\centering
		\includegraphics[scale=0.6]{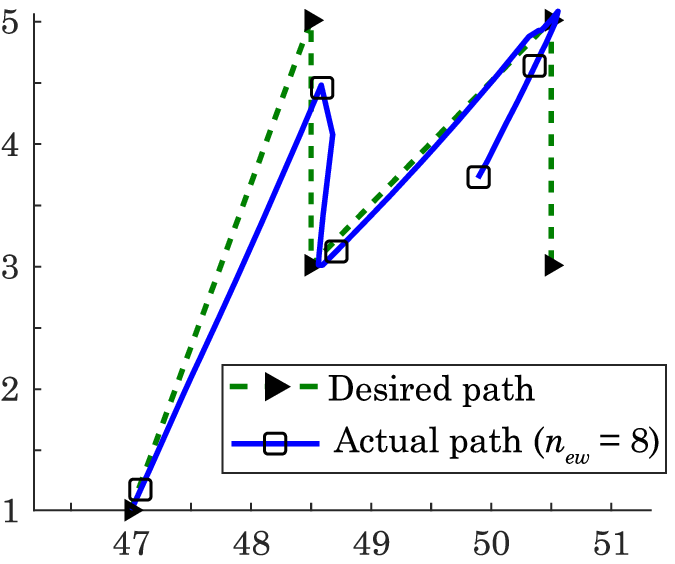}
		\caption{$ n_{ew} = 8 $}
		\label{fig:new_8_nel_20}
	\end{subfigure}
	\caption{Effect of various parameters on performance of the 3-kink CCM. (\subref{fig:Friction_mu_0_001}-\subref{fig:Friction_mu_0_1}) compares the desired and actual paths for different coefficients of friction i.e., $ \mu = 0.001, 0.01, 0.05 $ and 0.1. Required input force $ F $ is increased such that the entire path is traced. (\subref{fig:Wear_mu_0.05_w_0.5}-\subref{fig:Wear_mu_0.05_w_10}) represent variation of output path with respect to different wear percentages ($ w $). Coefficient of friction $ \mu = 0.05$  for all four cases. Input force $ F $ is 30.35N for all cases. (\subref{fig:new_5_nel_20}-\subref{fig:new_8_nel_20}) represent comparison of desired and actual paths for different mesh densities.}\label{fig:Mesh_Density}
	\label{fig:Friction} \label{fig:Wear}
\end{figure}

\subsection{Effect of wear}
To demonstrate the effect of wear on performance of 3-kink switch, we model wear by uniformly shrinking regions of the CCM that come in contact with external surfaces of the key. This is achieved by reducing the width of each quadrilateral element of such regions by $w = 0.5\%, 1\%, 5\%, \mbox{ and } 10\%$. Here, we assume that the CCM wears out uniformly at all contact-sites and no wear is considered for rigid external contact surfaces. With the above assumptions and methodology, mechanism tracing the 3-kink path is analysed considering the coefficient of friction as $0.05$ with the aforementioned wear percentages at contact locations. Figs. \ref{fig:Wear}\subref{fig:Wear_mu_0.05_w_0.5}-\subref{fig:Wear_mu_0.05_w_10} show the effect of wear. Change in objective function value with respect to different wear percentages are given in Table \ref{Table:Wear}. The output paths deviate but not much from the desired suggesting that even when the contact regions of the mechanism wear out, the latter will still retain its functionality. Further, from Figs. \ref{fig:Wear_mu_0.05_w_5} and \ref{fig:Wear_mu_0.05_w_10} one observes an increase in length of the output path that is because of reduced stiffness of CCM members at contact-sites, due to wear. Contribution of wear at each contact-site to overall deviation (though less) of output path will be explored in future studies along with wear of external surfaces.

\begin{table}[h!]
	\centering
	\caption{Effect of wear on performance of CCMs}
	\label{Table:Wear}
	\begin{tabular}{ccc}\toprule
		Wear percentage ($ w $) & Objective value & Output path                             \\ \toprule
		0.5             & 41.7132        & Fig. \ref{fig:Wear_mu_0.05_w_0.5}                                 \\
		1               & 38.5558        & Fig. \ref{fig:Wear_mu_0.05_w_1}                                   \\
		5               & 39.744         & Fig. \ref{fig:Wear_mu_0.05_w_5}                                   \\
		10              & 46.1985        & Fig. \ref{fig:Wear_mu_0.05_w_10}					\\ \bottomrule                              
	\end{tabular}
\end{table}

\subsection{Computational time}
Time involved in generating an optimized CCM design comprises time consumed for each analysis and the time taken for the optimization process.

\subsubsection{Time for single analysis}
\begin{table}[h!]
	\centering
	\caption{Comparison of time taken for single analysis with respect to mesh size}
	\label{Table:Time_Mesh_Size}
	\begin{tabular}{@{}ccccccc@{}}
		\toprule
		Sr.No & new & nel & Total number of elements & Objective function value & Time taken (sec) & Path    \\ \midrule
		1     & 5   & 20  & 4060                     & 41.3980                  & 134              & Fig. \ref{fig:new_5_nel_20} \\
		2     & 6   & 20  & 4856                     & 41.9402                  & 158              & Fig. \ref{fig:new_6_nel_20} \\
		3     & 7   & 20  & 5740                     & 44.0356                   & 186              & Fig. \ref{fig:new_7_nel_20} \\
		4     & 8   & 20  & 6712                     & 45.047                   & 203              & Fig. \ref{fig:new_8_nel_20} \\ \bottomrule
	\end{tabular}
\end{table}

For the 3-kink CCM, typical clock-time for generation and analysis of a candidate design is approximately 120 seconds. Mesh generation, plotting, identifying contact surfaces as loops, specifying contact interactions amongst them and preparation of \texttt{ABAQUS\textsuperscript{TM}} input file takes approximately 10 seconds on a desktop computer equipped with Intel(R) i5-6400 CPU @ 2.71GHz processor and 16 GB RAM. The remaining time goes into analysis and extraction of nodal displacements. Considering large number of quadrilateral elements and nodes the analysis time is justified. Further, considering self and mutual contact modes and book keeping of contact related information, based on our experience,  this time cannot be avoided even if one develops an in-house code. A comparison of time taken for single analysis with respect to number of elements and nodes is presented in Table \ref{Table:Time_Mesh_Size}. Also, implementation of friction is possible which may take marginally additional time (presented in Sec. \ref{sec:friction}). A comparative study is performed to demonstrate variation of output path with respect to mesh density and the same is presented in Figs. \ref{fig:Mesh_Density}\subref{fig:new_5_nel_20}-\subref{fig:new_8_nel_20}. Number of elements along width of member ($ n_{ew} $) is varied to modify mesh density. From Figs. \ref{fig:Mesh_Density}\subref{fig:new_5_nel_20}-\subref{fig:new_8_nel_20}, one observes that optimum mesh density needs to be chosen for better results. Slender elements may be formed if number of elements along width are not in proportion to number of elements along length, which may leads to more deviation in output path as given in Fig. \ref{fig:new_8_nel_20}. Time reduction for single analysis is possible but perhaps at the cost of accuracy.

\subsubsection{Search algorithm and time taken} 
We chose zero order Hill Climber search which is expected to yield a desired CCM requiring significantly large number of function evaluations compared to a gradient search and hence, it is computationally expensive. Hill Climber search is chosen to cater to discrete variables responsible not only to generate external surfaces but also to identify contact surface types (elliptical, circular and rectangular). Also, discrete variables are used in determining the continuum topology. In addition, final/intermediate designs are also evaluated for feasibility of contact analysis by checking whether external surfaces intersect with skeleton of continuum members or whether the fleshed out members. In the first case, discrete variables are modified to remove the intersecting members while in the second case the same is done to remove external surfaces. The overall goal is to consider as many potential CCM designs for analysis. The aforementioned may not be possible with gradient search. In addition, a gradient search will not proceed in case the analysis for an intermediate design does not converge. With zero order searches, such designs can be penalized and ignored, facilitating smooth/uninterrupted search.

\section{Closure}
\label{sec:conclusion}
A methodology to systematically design and synthesize CCMs that trace non-smooth paths with multiple kinks using self and mutual contact with external surfaces of varying shapes is presented. A novel method of identifying contact surfaces as closed loops is conceived. Geometrically nonlinear finite element analysis is performed using \texttt{ABAQUS\textsuperscript{TM}}. Since, it is felt that a gradient search cannot be employed, a zero order Hill Climber search is adopted to easily penalize incomplete/non-convergent intermediate designs. Fourier shape descriptors are employed to compare desired and actual paths. A three-kink CCM switch is designed using the proposed methodology. Functionality of CCM switch with different wrong key combinations is studied. It is observed that lock does not open with any of these candidate wrong keys, however, in very few cases the alarm does not get activated.  We further perform a detailed study on effect of friction and wear on performance of CCM switch. With friction at contact-sites, more input force is required to get the same output path. This additional force requirement increases significantly with  increase in friction coefficient. Thus, use of smooth and polished external surfaces is recommended to minimize friction. Wear at contact-sites is modeled by shrinking the quadrilateral elements of the continuum at those sites. It is observed that wear does not have much impact on the desired deformation of the CCM. Further, mesh independence is demonstrated. It is suggested that the number of elements along width and length directions should be proportional to member dimensions to avoid elements with high aspect ratio, and also for efficient analysis. In future, we intend to explore more features/applications of CCMs, including static balancing, with generically shaped external surfaces, either rigid or deformable.

\section*{Acknowledgment}
\addcontentsline{toc}{section}{Acknowledgment}
First author acknowledges the support received from Knowledge Incubation for TEQIP (KIT) Office, IIT Kanpur and its staff, during his visit to IIT Kanpur. He also acknowledges the help received from fellow CARS lab members, Mr. Shyam Sundar Nishad, Mr. Nikhil and Mr. Vitthal during the initial stages of preparation of this manuscript.

\bibliography{References}
\bibliographystyle{ieeetr}

\end{document}